\documentclass[aps,floatfix]{revtex4-2}
\usepackage{hyperref,hypcap,ae,tikz}
\usepackage{graphicx}
\usepackage{bm}
\usepackage{amsmath}
\usepackage[T1]{fontenc}
\usepackage{textcomp}
\usepackage{xcolor}
\usepackage{lmodern}
\usepackage[caption=false]{subfig}
\usepackage{mathtools}

\hypersetup{
	colorlinks=true,
	citecolor=blue,
    linkcolor=magenta,
	}
\begin{document}

\title{{\bf  How appropriate are the gravitational entropy proposals for traversable wormholes? }}

\author{\bf Samarjit Chakraborty$^{1,a}$}
\author{Sarbari Guha$^{1,b}$}
\author{Rituparno Goswami$^{2,c}$}

\affiliation{\bf $^1$ Department of Physics, St.Xavier's College (Autonomous), Kolkata 700016, India \\
$^2$ Astrophysics and Cosmology Research Unit, School of Mathematics, Statistics and Computer Science,
University of KwaZulu-Natal, Private Bag X54001, Durban 4000, South Africa \\
$^{a}$ samarjitxxx@gmail.com \\
$^{b}$ srayguha@yahoo.com; guha@sxccal.edu \\
$^c$ vitasta9@gmail.com; goswami@ukzn.ac.za}

\maketitle
\section*{Abstract}
In this paper we have examined the validity of some proposed definitions of gravitational entropy (GE) in the context of traversable wormhole solutions of the Einstein field equations. Here we have adopted two different proposals of GE and checked for their applicability in the case of these wormholes. The first one is the phenomenological approach proposed by Rudjord et al \cite{entropy1} and expanded by Romero et al in \cite{entropy2}, which is a purely geometric method of measuring gravitational entropy. The latter one is the Clifton-Ellis-Tavakol (CET) proposal \cite{CET} for the gravitational entropy which arises in relativistic thermodynamics, and is based on the Bel-Robinson tensor, that represents the effective super-energy-momentum tensor of free gravitational fields. Considering some of the Lorentzian traversable wormholes along with the Brill solution for NUT wormholes and the AdS wormholes, we have evaluated the gravitational entropy for these systems. Incidentally, the application of the CET proposal can provide unique gravitational entropies for spacetimes of Petrov type D and N only, whereas the geometric method can be applied to almost every kind of spacetime, although it has no relation with thermodynamics. For any traversable wormhole to be physically realistic, it should have a viable GE. We found that the GE proposals do give us a consistent measure of GE in several of them. This means that the existence of a viable gravitational entropy strictly depends on its definition.

\bigskip

KEYWORDS: Gravitational entropy, Traversable wormhole.

\section{Introduction}
Gravitational entropy is the entropy measure reflecting the degrees of freedom associated with the free gravitational field. In any physical process involving gravity, the clumping of matter (structure formation) or the intensity of gravitational field in a local region of spacetime can be measured in terms of this quantity. Historically, this idea was proposed to justify the low entropy state of the initial universe, i.e., entropy is associated with the free gravitational field even during the time of big bang, so that a gravity dominated evolution of the universe does not violate the second law of thermodynamics. Another importance of gravitational entropy is that it puts the black hole entropy in the proper context, making it the special case of the gravitational entropy of free gravitational fields. Black hole (BH) entropy has a special position in physics because it is the only entropy which is proportional to the area of the gravitating object, unlike other thermodynamic entropies which are proportional to the volume.

Gravitational entropy proposals (GE proposals) can be studied in both local and global contexts. Locally, BH entropy (or the entropy of any astrophysical object) represents the immense concentration of entropy in a given region of spacetime. Thus gravity condenses matter leading to the increase in entropy of the universe, and BH entropy is the ultimate result of that process. Global studies on the evolution of the universe also indicate that the GE is monotonically increasing and well-behaved near the initial singularity. The study of GE is also important in thermodynamics. In gravitational theories, geometry and energy are interrelated, and both kinds of measures (geometric and thermodynamic) are available for GE. It not only gives us an idea of the nature of some specific geometry but also encapsulates the overall energetics of that region. The study of GE also tells us how matter and free gravitational fields behave in a particular region or in an overall fashion. Thus the concept of GE is still developing, as pointed out in \cite{CET}.

Roger Penrose’s Weyl curvature hypothesis (WCH) \cite{Penrose1} was the first effort to understand GE in a formal way. He argued that entropy can be assigned to the free gravitational field, and the Weyl curvature serves as a measure of it. It was assumed that the universe began from a singular state where the Weyl component was much smaller compared to the Ricci component. To this end, the FLRW models provide us an approximate description of the  early phase of the universe. The Weyl curvature was zero at early times for the FLRW case, but is large in the Schwarzschild-like spacetimes, which represents the geometry of a spherically symmetric star or a black hole formed during the later phases of evolution, thereby indicating the validity of the Weyl curvature hypothesis, as free gravitational entropy is larger in strongly gravitating systems than in flat spacetimes. This is the kind of behavior that we expect from a description of GE, i.e., it should increase throughout the history of the universe, in agreement with the second law of thermodynamics \cite{Penrose2,Bolejko}. Even then, there is still doubt about the definition of gravitational entropy in a way similar to thermodynamic entropy, which may be applied to all gravitating systems \cite{CET}.

In $ 2008 $, Rudjord, Grøn and Hervik \cite{entropy1} pointed out that the Weyl scalar is not a good measure of GE. The same is true for the ratio of the Weyl scalar squared to the squared Ricci tensor. They explicitly showed that their proposal of GE, i.e., the ratio of the Weyl scalar to the Kretschmann scalar, serves as a good measure and reproduces the Hawking-Bekenstein entropy for black holes \cite{SWH1,Bekenstein}. Later in $ 2012 $, Romero, Thomas and Pérez \cite{entropy2} applied it to other systems of black holes and wormholes, validating and extending the proposal of Rudjord et al. In $2013$, Clifton et al. \cite{CET} provided a measure of gravitational entropy based on the square root of the Bel-Robinson tensor, an idea motivated by thermodynamic considerations, which has a natural interpretation as the effective super-energy-momentum tensor of free gravitational fields. This is the so-called Clifton-Ellis-Tavakol (CET) proposal of gravitational entropy. But this definition is only valid for General Relativity (GR), where the Bel-Robinson tensor can be defined in such a way. Among the above two measures of gravitational entropy, the one due to Rudjord et al. and Romero et al. represents a geometrical measure, whereas the CET proposal gives us a thermodynamic measure of GE.

Sussman \cite{suss1} introduced a weighed scalar average formalism (the ``q-average'' formalism) for the study  of spherically symmetric LTB dust models and considered the application of this formalism to the definition of gravitational entropy functional proposed by Hosoya et al. (HB proposal) \cite{HB}. Subsequently, Sussman and Larena \cite{suss2} analyzed the generic LTB dust models to probe the CET proposal and the HB proposal (also a variant of the HB proposal), suggesting that the notion of gravitational entropy is theoretically robust, and can also be applied to many other generic spacetimes. The same authors studied the evolution of the CET gravitational entropy for the local expanding cosmic CDM voids using a nonperturbative approach \cite{suss3}.

Gravitational entropy can also be computed for objects like black holes and wormholes. Wormholes are shortcuts between two spacetime points. In most cases, these holes are unstable or causally disconnected, i.e., observers encounter closed timelike geodesics while travelling through them. Here we are interested in the traversable wormholes as they are physically more interesting because observers do not violate any causality while crossing them, there are no horizons, nor any curvature singularities giving rise to exotic physical scenarios \cite{visser1}. The study of gravitational entropy in the case of the traversable wormholes is important because it gives us a thermodynamic perspective of the physical reality of these solutions. If the gravitational entropy is well behaved, then these solutions are thermodynamically robust. Otherwise it may be thermodynamically unstable, and thus traversability may not be guaranteed. We now present a brief review of the important solutions of traversable wormholes.

The idea of traversable wormhole was first proposed by H. G. Ellis in 1973 \cite{ellis}, who showed that a coupling of geometry with a scalar field $ \phi $ produces a static, spherically symmetric, geodesically complete, horizonless space-time manifold with a topological hole (termed as a \emph{drainhole}) in its center. In 2005, Das and Kar \cite{ellis2} showed that the Ellis wormhole can also be obtained using tachyon matter as the source with a positive cosmological constant in $3+1$ dimensions. The Ellis wormhole has been studied by several authors \cite{ellis3,ellis4,ellis5} mainly in the field of deflection of light by massive objects, gravitational microlensing and wormhole shadows \cite{ellis1}. In \cite{ellis6} it was shown that when the throat is set into rotation, the static wormhole evolves into a rotating 4-dimensional Ellis wormhole supported by phantom scalar field. Recently, the Ellis wormhole without a phantom scalar field has been demonstrated \cite{ellis7} in $ 3+1 $ dimensional Einstein-scalar-Gauss-Bonnet theory (EsGB) in electrovacuum. By nonminimally coupling the phantom scalar field with the Maxwell field in \cite{ellis8}, the authors obtained charged Ellis wormhole and black hole solutions in the Einstein-Maxwell-scalar theory.

Independently, Bronnikov in 1973 proposed the same idea in scalar tensor theories on vacuum and electrovacuum spherically symmetric static solutions \cite{bro}. Subsequently Morris and Thorne also discussed the traversable wormholes in their 1988 paper \cite{mt}. Matt Visser in 1989 \cite{visser2} discussed some examples of traversable wormholes. In 1995, Cramer et.al. \cite{cramer} proposed another kind of traversable wormhole with negative mass cosmic strings which might have occurred in the early universe. In 1954, Papapetrou proposed the exponential spherically symmetric metric induced by scalar and antiscalar background fields \cite{Papapetrou} which represents a counterpart to the Schwarzschild black hole. Recently \cite{exp2}, it was shown that this metric has its origin within a wide class of scalar and antiscalar solutions of the Einstein equations parameterized by scalar charge. The exact rotational generalization of the antiscalar Papapetrou spacetime as a viable alternative to the Kerr black hole has been studied in \cite{exp4}. The Darmour-Solodukhin wormhole was proposed in $2007$ by T. Darmour and S. N. Solodukhin \cite{DS1}. Recently Matyjasek \cite{DS2} have demonstrated that for scalar fields there is a parameter space which allows this wormhole to be a traversable one.

The solution of the Einstein-Maxwell system of equations was found by Brill \cite{brill} in $1964$. Recently, the thermodynamics of the Taub-NUT solution has been studied in the Euclidean sector by imposing the condition for the absence of Misner strings \cite{NUT1}. More work have been done recently by several authors \cite{NUT2,NUT3,NUT4,NUT6,NUT7} on the various thermodynamic issues. Clement et.al. \cite{NUT5} investigated the electrically charged particle motions for such a metric. More recently \cite{NUT8}, nonlinear extensions of gravitating dyon-like NUT wormholes have been studied. A recent work by J. Podolský on accelerating NUT blackholes \cite{pod} shows that accelerating NUT blackholes act as a throat of maximal curvature connecting our universe $(r > 0)$ with the second  parallel universe in the region $(r < 0)$.

Richarte and Simeone constructed thin-shell Lorentzian wormholes with spherical symmetry in 5-dimensional Einstein-Gauss-Bonnet theory. For certain values of the parameters, these wormholes could be supported by ordinary matter \cite{WH0}. Clement et. al \cite{WH1} showed that traversable wormholes can exist without exotic matter but with a NUT parameter, and there is no causality violation in such cases. Beato et. al. \cite{WH2} found that the self-gravitating, analytic and globally regular Skyrmion solution of the Einstein–Skyrme system in presence of a cosmological constant has a non-trivial byproduct representing traversable AdS wormholes with NUT parameter, in which the only ``exotic matter'' required for their construction is a negative cosmological constant. Carvente et.al. \cite{lWH} studied traversable $ \ell $-wormholes supported by ghost scalar fields. Lima et. al. \cite{lima} have recently calculated the gravitational entropy for wormholes with exotic matter and in galactic halos.

G. Horowitz et. al. in 2019 discussed the nucleation process of traversable wormholes through a nonperturbative process in quantum gravity \cite{WHn}. In the same year Mattingly et.al. \cite{WHi} determined the curvature invariants of Lorentzian traversable wormholes. In \cite{torsion} the authors have shown that the violation of the null energy condition by matter, required by the traversable wormholes, can be avoided in spacetimes with torsion. Sebastiani et. al. \cite{wkb} proposed a unified classical approach for the studying idealized gravitational compact objects like wormholes (WHs) and horizonless stars by using the characteristic echoes generated in the ringdown phase.
Therefore, the study of traversable WHs becomes important in the context of GE, as more and
more of such studies have revealed that these systems may exist as possible astrophysical objects. If traversable WHs do exist, then the different proposals of GE must be tested on them to see whether they exhibit a viable GE.

There have been many definitions of black hole entropy and wormhole entropy using quantized theories of gravity, such as string theory and loop quantum gravity. However, in this paper we have addressed the problem from two different perspectives: the first one is the phenomenological approach proposed in \cite{entropy1} and expanded in \cite{entropy2}, in which the Weyl curvature hypothesis was tested against the expressions for the entropy of cosmological models and black holes. Comprehensive study of the various proposals of gravitational entropy in the case of various traversable wormholes is not available in literature. This has prompted us to analyze the gravitational entropy of wormholes in terms of the second perspective: the CET proposal, which yields us a pure thermodynamic measure of GE.

In our previous works on the accelerating black holes \cite{GC} and the cosmological models \cite{CGG}, we examined the validity of the Weyl scalar proposal of GE (proposed by Rudjord et al.), and the CET proposal of GE, respectively, for two different types of systems. In this paper we are extending our previous studies on gravitational entropy to the case of traversable wormholes using these two proposals. As the Weyl scalar proposal is based on a purely geometric approach, whereas the CET proposal includes the details of relativistic thermodynamics, it is interesting to study them side by side on the same system to see how these proposals differ from each other when applied to a specific spacetime geometry. A similar study was done by Pérez et al \cite{kerr} in the context of Kerr black holes. One purpose of this current work is also to see how our results for traversable wormholes compare with the results found in \cite{kerr}. These comparisons will provide us a better understanding of these different estimators of GE.

In this paper, we have examined various traversable wormholes from the simplest to the more involved ones. The Ellis wormhole is the simplest case. It is a zero mass traversable WH which connects two asymptotically flat regions at its throat. There have been many proposals for the energy source of such WHs making it not only an ideal toy model to study, but also having rich physical content. In the same spirit we have considered the exponential metric WH, which is a much more general spherically symmetric spacetime, and is traversable at the throat. There are WHs which can also mimic BHs, and the simplest case of such a WH is the Darmour-Solodukhin (DS) WH. This is of great physical interest as it is not only traversable, but also because it mimics the Schwarzschild BH to an outside observer, for all practical purposes. In the spherically symmetric static cases, both the exponential metric WH and the DS WH possess rich mathematical and physical structure. In order to examine the consequence of charge present in the WH system on the GE of that system, we considered the Maldacena ansatz, which connects two oppositely charged BHs. As for the stationary cases, we considered the Brill NUT WH (where both the magnetic and electric charges are present), and is an extension of the Reissner-Nordström solution with a Newman–Unti–Tamburino (NUT) parameter. This would enable us to study the behaviour of GE in Einstein-Maxwell systems, where the NUT parameter controls the WH neck. Extending this to cosmological settings, we consider an AdS NUT WH in the presence of a negative cosmological constant, so as to study the effect of $ \Lambda $ on the gravitational entropy of NUT WH. We have considered these widely different traversable WHs, in order to study the behaviour of GE explicitly in such scenarios, and to determine whether the GE proposals considered by us are physically viable or not.

Our paper is organized as follows: In Section II we have explained the gravitational entropy proposal given by Rudjord et al \cite{entropy1} and expanded by Romero et al \cite{entropy2}, and the CET proposal of gravitational entropy \cite{CET}. In the next section we have analysed the gravitational entropy for these systems. In section IV we have tested the validity of the Tolman law in the case of static spherically symmetric WHs, and subsequently presented the summary of our work and the concluding remarks in sections V and VI. Finally, in the appendix we have also included a brief analysis of the traversable AdS wormhole and followed it up for a general wormhole ansatz recently proposed by Maldacena.

\section{Gravitational Entropy}
In this section we will describe briefly the two proposals of gravitational entropy, namely the Weyl scalar proposal proposed by Rudjord et al. \cite{entropy1} and extended by Romero et al. \cite{entropy2}, and the CET proposal given by Clifton et al. \cite{CET}. As chronologically the Rudjord et al. proposal came first, and then the CET proposal, we will be following this sequence in our description and subsequent analysis.

\subsection{\textbf{{Weyl scalar proposals}}}

Here we provide a brief description of the proposal given in \cite{entropy1} for the determination of gravitational entropy.
The entropy on a surface is described by the surface integral
\begin{equation}
S_{\sigma}=k_{s}\int_{\sigma}\boldsymbol{\Psi}.d\sigma ,
\end{equation}
where $ \sigma $ denotes the surface and the vector field $\boldsymbol{\Psi}$ is given by \cite{entropy1}:
\begin{equation}
\boldsymbol{\Psi}=P_{i} \boldsymbol{e_{r}},
\end{equation}
with $ e_{r} $ as a unit radial vector. Here $P_{i}$ represents either $ P_{1} $ or $ P_{2} $, which are described below. The scalar $ P_{1} $ is defined in terms of the Weyl scalar ($ W $) and the Krestchmann scalar ($ K $) in the form \cite{entropy1}:
\begin{equation}\label{P_sq}
P_{1}^2=\dfrac{W}{K}=\dfrac{C_{abcd}C^{abcd}}{R_{abcd}R^{abcd}},
\end{equation}
where the Weyl tensor in $ n $ dimensions is given by \cite{Chandra}
\begin{equation}
C_{\alpha\beta\gamma\delta}= R_{\alpha\beta\gamma\delta}+\dfrac{2}{(n-2)}(g_{\alpha[\gamma})R_{\delta]\beta}-g_{\beta[\gamma})R_{\delta]\alpha})
+\dfrac{2}{(n-1)(n-2)}Rg_{\alpha[\gamma}g_{\delta]\beta}.
\end{equation}
Equation \eqref{P_sq} is a purely geometric measure of GE, and hence it nicely encompasses the curvature dynamics. Here, the gravitational entropy is evaluated by doing computations in a 3-space. The spatial metric $ h_{ab} $ is defined as:
\begin{equation}
h_{ij}=g_{ij}-\dfrac{g_{i0}g_{j0}}{g_{00}},
\end{equation}
where $ g_{\mu\nu} $ is the corresponding 4-dimensional space-time metric, and Latin indices denote spatial components, $i, j = 1, 2, 3$. So the infinitesimal surface element is given by:
\begin{equation}
d\sigma=\dfrac{\sqrt{h}}{\sqrt{h_{rr}}}d\theta d\phi.
\end{equation}
Since wormholes does not have any horizons, it is preferable to switch into entropy density, $s$. We imagine an enclosed hypersurface, and apply Gauss's divergence theorem to find the entropy density \cite{entropy1} as the following:
\begin{equation}
s=k_{s}|\nabla.\Psi|.
\end{equation}
If we only consider the radial contribution of the vector $ \Psi $ then the gravitational entropy density becomes
\begin{equation}\label{sd1}
s=k_{s}|\nabla.\Psi|=\dfrac{k_{s}}{\sqrt{-g}}\left\vert\dfrac{\partial}{\partial r}(\sqrt{-g}P_{i})\right\vert.
\end{equation}
This is useful in the spacetimes with spherical symmetry.

We have also discussed the possibility of having an angular component in the vector field $\boldsymbol{ \Psi} $ for axisymmetric spacetimes as proposed in \cite{entropy2}. Using this modified definition of $ \boldsymbol{\Psi} $, we can calculate the gravitational entropy density for axisymmetric space-times, which is given by the following expression:
\begin{equation}\label{sd2}
s=k_{s}|\nabla.\Psi|=\dfrac{k_{s}}{\sqrt{-g}}\left\vert\left(\dfrac{\partial}{\partial r}(\sqrt{-g}P_{i})+\dfrac{\partial}{\partial \theta}(\sqrt{-g}P_{i}) \right)\right\vert.
\end{equation}
Therefore the $ P_{1} $, is not the only case that we have considered. We have also used the measure proposed in \cite{entropy2} for the expression of $ P_{i} $ in the case of metrics having nonzero $ g_{t\phi} $ component, which is given below:
\begin{equation}\label{p2}
P_{2}=C_{abcd}C^{abcd}.
\end{equation}
Using this definition of $ P_{2} $ we have calculated the gravitational entropy density for the relevant wormholes.
It may be noted that the pure Weyl square proposal (in Eq. \eqref{p2}) fails at isotropic singularities, and
cannot handle the decaying and growing perturbation modes. As these quantities are purely geometric in nature that incorporate the connection of the Weyl tensor with the free gravitational field, they do not provide a theoretical connection with thermodynamics or Information theory. Further, as the ratio in Eq. \eqref{P_sq} is a dimensionless scalar, it cannot be related to the Hawking-Bekenstein entropy. Moreover, these geometric proposals are frame-independent, and hence have no connection with the worldlines of physical fluids. While $ P_{1} $ can address the above objections, but it does not seem to give the correct sense of time for a radiating source (see \cite{CET} and references therein).

\subsection{\textbf{Clifton-Ellis-Tavakol (CET) proposal}}
For the static spherically symmetric WHs we have also used the CET proposal \cite{CET} to examine the behaviour of GE for these systems. To establish the validity of this proposal, the CET paper has shown that it not only reproduces the Hawking-Bekenstein entropy for BHs, but also its entropy production rate, $ \dot{s}_{grav} $, is always non-negative.
This proposal begins with the construction of the second order symmetric traceless tensor $ t_{ab} $ which is obtained from the algebraic
``square root'' of the fourth order Bel-Robinson tensor $T_{abcd}$, because $T_{abcd}$ is the only totally symmetric traceless tensor that can be constructed out of the conformal Weyl tensor $ C_{abcd} $, and secondly, $T_{abcd}$ is fourth order with dimensions as $ L^{-4} $ making its ``square root'' necessary. This $ t_{ab} $ helps us to derive the ``effective'' or ``super energy–momentum tensor'' $ \mathcal{T}_{ab} $ of the free gravitational field. We can also compute other variables like gravitational energy density $ \rho_{grav} $, gravitational pressure $ p_{grav} $, anisotropic stresses $ \Pi^{ab}_{grav} $, and heat flux $ q^{a}_{grav} $ by contracting with the matter 4-velocity $ u^{a} $ and projector tensor $h_{ab} = u_{a}u_{b} + g_{ab}$. Subsequently, by analogy with the standard laws of fluid thermodynamics applied on the quantities associated with $\mathcal{T}_{ab}$, the notion of gravitational entropy emerges clearly. In \cite{CET}, the authors have considered two types of gravitational fields: the ``Coulomb-like'' (Petrov type D) and the ``wave-like'' (Petrov type N) fields for which $ \mathcal{T}_{ab} $ reduces to expressions involving the Newman–Penrose conformal invariants $ \Psi_{2} $ and $ \Psi_{4} $. We will be restraining our discussions to Coulomb-like fields, and therefore will only discuss the Petrov type D case. For this case, CET derived the following $ \mathcal{T}_{ab} $ and the associated fluxes:\\
\begin{align}\label{effT}
&\frac{{\mathcal{T}}^{ab}}{8\pi}=\alpha|\Psi_2|\left[ x^{a}x^{b}+y^{a}y^{b}-2\left(z^{a}z^{b}-u^{a}u^{b}\right)\right]=\rho_{grav} u^{a}u^{b} + p_{grav} h^{ab}+2q^{(a}_{grav} u^{b)}+\Pi_{grav}^{ab},\nonumber\\
& \left.\right. \\
& 8\pi\rho_{grav}=2\alpha|\Psi_2|,\quad p_{grav}= q_{grav}=0,\quad 8\pi\Pi_{grav}^{ab}=\frac{\alpha|\Psi_2|}{2}(x^{a}x^{b}+y^{a}y^{b}-z^{a}z^{b}+u^{a}u^{b}).\nonumber
\end{align}
Further, the gravito-electromagnetic properties of the Weyl tensor, and the 1+3 decomposition of the equations is used to express the gravitational ``Super energy density'' function $w$ as follows:
\begin{equation}
w=T_{abcd}u^{a}u^{b}u^{c}u^{d}=\dfrac{1}{4}\left(E_{a}^{b}E_{b}^{a}+ H_{a}^{b}H_{b}^{a}\right).
\end{equation}
Here, $ \alpha $ is a positive constant which provides appropriate physical units, and $\left[u^{a}, x^{a}, y^{a}, z^{a}\right]$
is an orthonormal tetrad. The quantities $ E_{ab}$, and $ H_{ab} $ are the electric and magnetic parts of the Weyl tensor $ C_{abcd} $ respectively.
For the mathematical computations, the complex null tetrad is defined as the following:
\begin{equation}
\label{nullt}
m^a = \frac{1}{\sqrt{2}} \left( x^a - i y^a \right), \quad
l^a = \frac{1}{\sqrt{2}} \left( u^a - z^a \right), \quad {\rm and} \quad
k^a = \frac{1}{\sqrt{2}} \left( u^a + z^a \right),
\end{equation}
where $x^a$, $y^a$ and $z^a$ are spacelike unit vectors, which constitute an orthonormal basis together with $u^a$. Using these, the entire metric can be rewritten in terms of these tetrads: $g_{ab} =2 m_{(a} \bar{m}_{b)} - 2 k_{(a} l_{b)}$, with $l^a$ and $k^a$ being aligned with the principal null directions. Therefore in this scheme of the free gravitational field \cite{CET}, the effective gravitational energy density can be written as:
\begin{equation}\label{Psi2}
8\pi\rho_{grav}=2\alpha\sqrt{\dfrac{2w}{3}}, \quad |\Psi_{2}|=\sqrt{\dfrac{2w}{3}},
\end{equation}
with  $\rho_{\rm grav} \geq 0$. Here $ \Psi_{2} $ is the nonzero Weyl scalar component for Petrov type D spacetimes.
In the CET paper \cite{CET}, the authors by analogy with the off-equilibrium Gibbs equation, obtained the following expression for the entropy production in the presence of perfect fluid matter field:
\begin{equation}
 T_{grav}\dot{s}_{grav} = (\rho_{grav} v)^{\cdot}=-v\sigma_{ab}\left[\Pi^{ab}_{grav}+\frac{4\pi(\rho+p)}{3\alpha|\Psi_2|}E^{ab}\right]. \label{gone}
\end{equation}
The independently defined local gravitational temperature at any point in spacetime was given by the following expression:
\begin{equation}
T_{grav}=\dfrac{|u_{a;b}l^{a}k^{b}|}{\pi}=\dfrac{|\dot{u_{a}}z^{a}+H+\sigma_{ab}z^{a}z^{b}|}{2\pi},
\end{equation}
where $z^{a}$ is a spacelike unit vector aligned with the Weyl principal tetrad, $ \sigma_{ab} $ is the shear tensor, $\dot{u_{a}}=u^{b}\nabla_{a}u_{b}$ is the $4-$acceleration, $ H=\dfrac{\Theta}{3} $ is the isotropic Hubble rate, and $\Theta\equiv\tilde{\nabla}_{c}u^{c}=h^{b}_{c}\nabla_{b}u^{c} $ is the isotropic expansion scalar.
As the above described variables depend on all the four spacetime variables, naturally the gravitational entropy one-form can be expressed as: $ ds_{grav}=\partial_{0}(s_{grav})dx^{0}+\partial_{i}(s_{grav})dx^{i} $. Consequently the Gibbs one-form becomes a system of four partial differential equations where $ T_{grav} $ acts an integrating factor, and these systems need not be integrable. On careful inspection it is evident that one needs to know about the physics of the microscopic theory of gravity to know the temperature of gravitational fields, and therefore, in \cite{CET} the authors considered the results of BH thermodynamics and quantum field theory in curved spacetime to propose the definition of $T_{grav}$. This definitionis not limited only to horizons, but is local, and can reproduce the Hawking temperature, the Unruh temperature and the temperature of de Sitter spacetime in the appropriate limits. Therefore, although the $ T_{grav} $ provided in \cite{CET} is an extra ingredient appearing along with their main proposal, and it is always possible to define a new temperature, still the concept of $ T_{grav} $ is rather well-motivated.

In the original CET paper \cite{CET}, the authors were interested in the gravitational entropy production, considering only the time derivative $\dot{s}_{grav}=u^{a}\partial_{a}s_{grav}$, along the worldlines of comoving observers. The full integrability of the Gibbs one-form have been discussed explicitly by Sussman and Larena in \cite{suss2,suss3}. The spherically symmetric static WHs that we have considered here, are the equilibrium cases where the condition $ \dot{s}_{grav}=u^{a}\partial_{a}s_{grav}=0 $ holds strictly, and therefore, for the one-form coordinate basis $ [dt,dr,d\theta,d\phi] $, the gravitational entropy one-form becomes
\begin{equation}
ds_{grav}=\partial_{r}(s_{grav})dr,
\end{equation}
and the Gibbs one-form reduces to a single ordinary differential equation:
\begin{equation}
T_{grav}\partial_{r}[s_{grav}]=\partial_{r}[\rho_{grav}],
\end{equation}
which leads to the rate of variation of the local piecewise gravitational entropy along the radial direction as:
\begin{equation}
\partial_{r}s_{grav}(r)=\dfrac{\rho_{grav}(r)v(r)}{T_{grav}(r)}.
\end{equation}
Finally, the variation of the local piecewise gravitational entropy of static spherically symmetric spacetimes $ s_{grav}(r) $ can be determined as:
\begin{equation}\label{totalen}
s_{grav}(r)=\int\dfrac{\rho_{grav}(r)v(r)}{T_{grav}(r)}dr,
\end{equation}
where the volume element is represented by $ v(r)=4\pi\sqrt{h} $, where $ h $ is the determinant of the projector tensor $ h_{ab} $. It may be further noted that the CET proposal is applicable only to Einstein's gravity. If we try to apply this proposal to different kinds of spacetimes, it is found that the CET proposal can provide unique gravitational entropy only for the Petrov type D and type N spacetimes. The algebraic decomposition of the Bel-Robinson tensor into a second order effective energy momentum tensor can be done for other Petrov type spacetimes also, but it is only in the Petrov type D and N spacetimes that the second order effective energy momentum tensor is unique \cite{bonsen}.

\section{Analysis of gravitational entropy}

In this section we will study some important traversable wormholes extending our previous works \cite{GC,CGG}. In each case we will be applying the two separate proposals of gravitational entropy if possible, and discuss their implications and limitations. We will also compare the corresponding results wherever possible.
We want to emphasize here that the energy momentum tensor for the  matter source supporting the WHs has no effect on the effective energy momentum tensor, $\mathcal{T}^{ab}$, for the free gravitational field given by CET and described in \eqref{effT}. It is clear from this equation that the geometric fluid for $\mathcal{T}^{ab}$ is not associated to the  matter source ($T^{ab}$) of WH.

\subsection{\textbf{Ellis wormhole}}
The Ellis WH obtained in 1973 \cite{ellis} is the first example of a non-singular wormhole solution. This is an exact solution of the Einstein-phantom scalar system with a scalar field having negative kinetic energy. Ellis used the negative kinetic energy term (‘phantom’) in the scalar field action to achieve the violation of energy condition, which is  necessary to support the wormhole. The metric of this wormhole is given by \cite{ellis}:
\begin{equation}\label{ellism}
ds^2=-dt^2+dr^2+(r^2+a^2)(d\theta^2+sin^2\theta d\phi^2),
\end{equation}
where $r=a $ is the throat radius of the wormhole.  The radial coordinate $ r $ runs from $ -\infty $ to $ +\infty $ to cover the entire wormhole geometry, where $ r=0 $ corresponds to the throat of the wormhole. This metric has no singularity, and the throat connects the two separate regions $ r\rightarrow +\infty $ and $ r\rightarrow -\infty $. The stress-energy tensor for this WH is: $-T^{tt}=-T^{rr}=T^{\theta\theta}=T^{\phi\phi}=\dfrac{a^2}{(a^2+r^2)^2} $, where the energy density is negative, which is only possible with exotic phantom matter.

\begin{enumerate}
\item \textbf{Weyl scalar proposal:} The spatial section of the Ellis wormhole is given by the following:
\begin{equation}
h_{ij}=diag\big[1,(r^2+a^2),(r^2+a^2)sin^2\theta \big].
\end{equation}\\
The determinant of the above mentioned matrix is given by $ h $:
$ h=(r^2+a^2)^2 sin^2\theta .$\\
The Weyl curvature scalar $ W $ and the Kretschmann curvature scalar $ K $ are obtained in the form:
$$ W= \frac{16}{3}\dfrac{a^4}{(a^2+r^2)^4}, \qquad \qquad \qquad K= \dfrac{12a^4}{(a^2+r^2)^4}.$$
Therefore the ratio of the two curvature scalars is given by:
$ P_1^2=\frac{W}{K}=\frac{4}{9} .$

\begin{figure}[ht]
    \centering
    \subfloat[Subfigure 1 list of figures text][]
        {
        \includegraphics[width=0.45\textwidth]{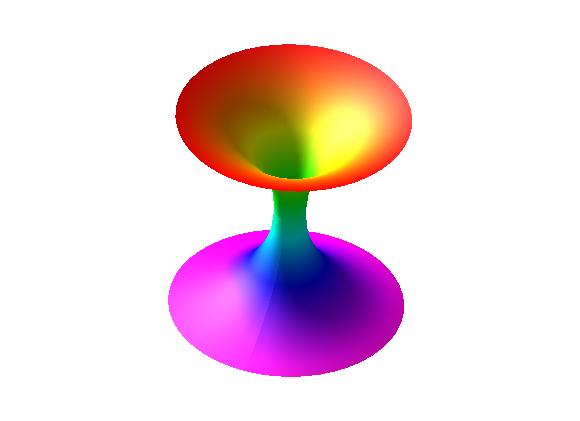}
        \label{fig:subfig1}
        }
    \subfloat[Subfigure 2 list of figures text][]
        {
        \includegraphics[width=0.45\textwidth]{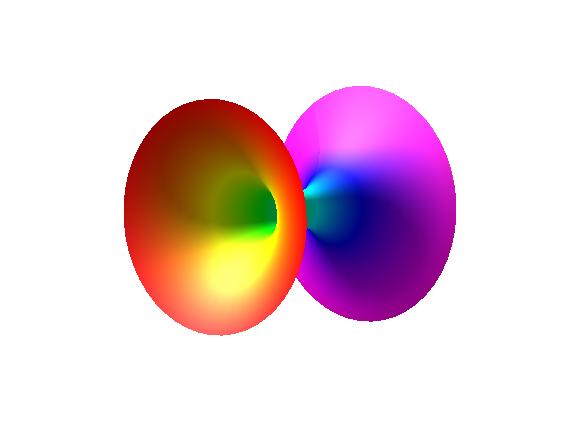}
        \label{fig:subfig2}
        }
    \caption{Embedding diagram of the Ellis wormhole where we have taken $ a=2 $.}
    \label{ellisem}
\end{figure}
The gravitational entropy density is obtained as
\begin{equation}\label{el}
s=k_{s}\vert\nabla.\Psi \vert= k_{s}\left\vert\frac{1}{\sqrt{h}}\dfrac{\partial}{\partial r}(\sqrt{h}\dfrac{P}{\sqrt{h_{rr}}}) \right\vert=\dfrac{4k_{s}}{3}\left\vert\dfrac{r}{(r^2+a^2)}\right\vert.
\end{equation}
Fig.\ref{ellisem} illustrates the embedding diagram of the Ellis wormhole for the values of parameters as mentioned.

\begin{figure}[ht]
    \centering
    \subfloat[Subfigure 1 list of figures text][]
    {\begin{minipage}[t][9cm][b]{.5\textwidth}
\centering
        {
        \includegraphics[width=0.97\textwidth]{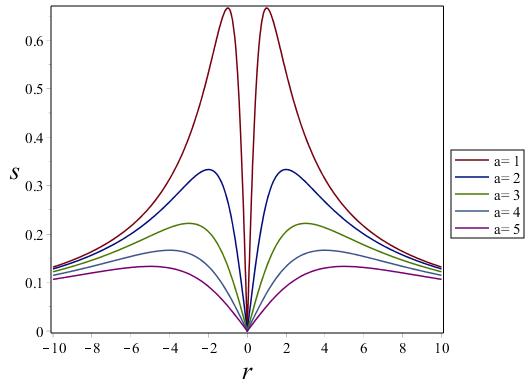}
        \label{fig:subfig1a}
        }
        \end{minipage}}
\subfloat[]{\begin{minipage}[t][9cm][b]{.5\textwidth}
\centering
\includegraphics[width=0.97\textwidth, height=0.2\textheight]{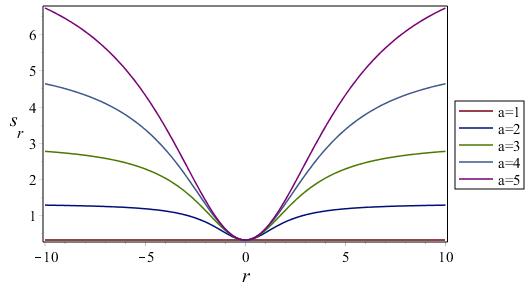}
\includegraphics[width=0.97\textwidth,height=0.2\textheight]{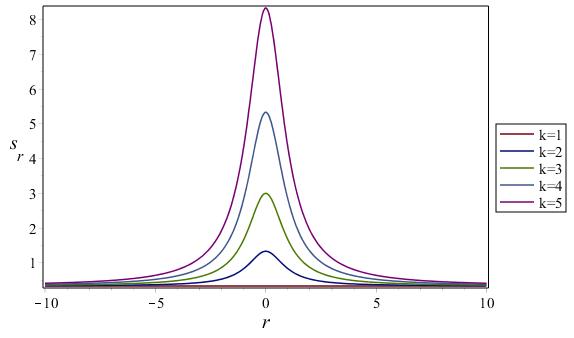}
\end{minipage}}
    \caption{(a) Variation of GE density $ s $ of Ellis wormhole with $ r $ for various throat lengths $ a $. (b) Rate of variation of CET gravitational entropy along the radial direction $\partial_{r}s_{grav}^{IF}\equiv s_{r} $ for different values of $ a $, where $ k=1 $, and for different values of $ k  $ where $ a=1 $, respectively.}
    \label{ellisentrop}
\end{figure}

In Fig.\ref{ellisentrop}(a) the gravitational entropy density derived in the equation \eqref{el} is depicted for different throat lengths. The maxima of gravitational entropy density lies at the wormhole throat radius, and the maxima of the entropy density is shifting to a lower value with the increasing throat radius. Further, the gravitational entropy density is always zero at the throat region of the wormhole as there is no central singularity in the metric of the Ellis wormhole. The finiteness of the gravitational entropy density also conforms to the traversability of this wormhole. It maybe noted that Romero et al. \cite{entropy2} evaluated the gravitational entropy density of another kind of traversable wormhole with exotic matter using the Weyl scalar proposal. They have also found that the gravitational entropy density is always zero at the throat of the wormhole irrespective of the value of the throat radius. The discussion of Fig.\ref{ellisentrop}(b) is given later.
\item \textbf{CET proposal:} Next we examine the CET proposal of gravitational entropy for the traversable Ellis wormhole. The gravitational epoch function $ w $ in this case is
\begin{equation}
w=T_{tttt}u^{t}u^{t}u^{t}u^{t}=\dfrac{a^4}{6(a^2+r^2)^4}.
\end{equation}
We have calculated all the Weyl scalars and we find that the only nonzero component is $\Psi_{2}  $ given by:
\begin{equation}
\Psi_{2}=-\frac{1}{3}\dfrac{a^2}{(a^2+r^2)^2}.
\end{equation}
Therefore the identity $ \vert \Psi_{2} \vert=\sqrt{\dfrac{2w}{3}} $ is satisfied in this Petrov type D spacetime. The energy density of the gravitational field is
\begin{equation}
\rho_{grav}=\frac{\alpha}{4\pi}\vert\Psi_{2}\vert=\frac{\alpha}{12\pi} \left\vert \dfrac{a^2}{(a^2+r^2)^2} \right\vert.
\end{equation}
As $ \dot u_{a}=0, \Theta=0, \sigma_{ab}=0  $ and $ \omega_{ab}=0 $, the gravitational temperature provided by CET in \cite{CET} emerges as
\begin{equation}
T_{grav}=0.
\end{equation}
The vanishing gravitational temperature is problematic because the GE will blow up in that case. As the CET GE is frame dependent, and we have considered observers only along a static frame, it would be very interesting to check for observers whose $ 4- $velocity is tangential to the worldlines of observers traversing the WH, to see whether it solves the problem of vanishing $ T_{grav} $. Let us consider the radial timelike geodesics on the  equatorial plane, and with the condition $ a^2>0 $, these geodesics can traverse the WH from one side to the other. Consequently the $ 4- $velocity in such a frame turns out to be $ u^{a}=(\sqrt{2},-1,0,0) $. For the sake of clarity, we state our orthonormal basis $ [u^{a},r^{a},\theta^{a},\phi^{a}] $ in the expressions below:
\begin{align}
 &u^{a}=(\sqrt{2},-1,0,0), &  &\theta^{a}=\left(0,0,\frac{1}{\sqrt{2}}\dfrac{1}{\sqrt{r^2+a^2}},\frac{1}{\sqrt{2}sin\theta}\dfrac{1}{\sqrt{r^2+a^2}}\right),\nonumber\\
 &r^{a}=(1,-\sqrt{2},0,0), &   &\phi^{a}=\left(0,0,\frac{1}{\sqrt{2}}\dfrac{1}{\sqrt{r^2+a^2}},-\frac{1}{\sqrt{2}sin\theta}\dfrac{1}{\sqrt{r^2+a^2}}\right).
\end{align}
Here $ u^{a} $ is a timelike unit vector, and the rest are spacelike unit vectors orthogonal to $ u^{a} $ and to each other. Consequently, we can form the complex null tetrad according to \eqref{nullt}. We obtain the gravitational energy density as $ \rho_{grav}=\dfrac{\alpha}{12\pi}\dfrac{a^2}{(a^2+r^2)^2} $.

Using the above mentioned $ 4- $velocity we found the congruence properties, which are: the $ 4- $acceleration $ \dot{u}_{a}=0 $, the expansion scalar $ \Theta(r)=\dfrac{-2r}{a^2+r^2} $, and the stress tensor $ \sigma_{ab} $ is given below:
\begin{equation}
\sigma_{00}=\dfrac{2r}{3(a^2+r^2)},\, \sigma_{01}=\sigma_{10}=\dfrac{2r\sqrt{2}}{3(a^2+r^2)},\,  \sigma_{11}=\dfrac{4r}{3(a^2+r^2)},\,\sigma_{22}=-\frac{r}{3},\,\sigma_{33}=-\dfrac{r\sin^2\theta}{3}.
\end{equation}
Therefore, the quantities required for the determination of $ T_{grav} $ becomes: $ \sigma_{ab}r^{a}r^{b}=\dfrac{2r}{3(a^2+r^2)} $,
and the isotropic Hubble rate is $ H=\frac{\Theta}{3}=-\dfrac{2r}{3(a^2+r^2)} $. Once again this yields $ T_{grav}=0 $.
This means that the ratio blows up, i.e., $\frac{\rho_{grav}}{T_{grav}}\rightarrow \infty $, if we consider the $ T_{grav} $ proposed by CET. Here the free gravitational energy density is finite locally but the gravitational temperature according to the CET proposal for the free gravitational field is vanishing, indicating the divergent local piecewise gravitational entropy. However, physically this is impossible. Although the CET proposal is independent of the definition of $T_{grav}$, at most we can say that the definition of temperature given in the CET paper is not suitable in this case. Since the CET definition of $T_{grav}$ is completely \emph{ad hoc}, and as $T_{grav}$ is simply an integrating factor of the Gibbs one-form, we can choose an ad hoc expression of $ T_{grav} $ in order to integrate the expression for GE. Now, if we want a finite entropy density at the throat and a non negative function of $ r $, then let us define the simplest integrating factor as $ T^{IF}_{grav}\sim\dfrac{1}{k^2+r^2}$, where $ k $ is a constant. It is to be noted that this temperature is for pure mathematical convenience and has no physical standing, whereas the temperature $T_{grav}$ given by CET was very robust since it could reproduce the Hawking-Bekenstein temperature for BH. This \textbf{\emph{temperature function}} $ T^{IF}_{grav}$ actually replicates the CET definition far away from the wormhole, at large values of ``r'', whereas near the wormhole it gives a finite non zero value. We will restrict our calculations to the equatorial plane ($ \sin\theta=1 $). The function $ T^{IF}_{grav}$ gives us the rate of variation of GE along the radial direction as
\begin{equation}
\partial_{r}s_{grav}^{IF}=\dfrac{\rho_{grav}v}{T^{IF}_{grav}}\sim\bigg\vert\dfrac{\alpha}{3}\dfrac{a^2 (k^2+r^2)}{(a^2+r^2)}\bigg\vert .
\end{equation}
In Fig.\ref{ellisentrop}(b) the rate of radial variation of CET GE, $\partial_{r}s_{grav}^{IF} $, is shown as a function of $ r $. Unlike the Weyl scalar proposal, here the entropy density is not always zero at the throat. The local piecewise CET gravitational entropy can be obtained by integrating over the radial coordinate as the following:
\begin{equation}
s_{grav}^{IF}(r)=\int_{0}^{r} \dfrac{\rho_{grav}v}{T^{IF}_{grav}}dr\sim\bigg\vert\dfrac{\alpha a}{3}\left((a^2-k^2)\arctan\left(\frac{r}{a}\right)-ar\right)\bigg\vert.
\end{equation}
Both the radial rate of variation of CET gravitational entropy, $ \partial_{r}s_{grav}^{IF} $, and the local piecewise GE, $ S_{grav}^{IF}(r) $, vanish at the throat for $ k=0 $, and for $ k\neq 0 $ the rate of variation of GE along the radial direction at the throat depends on the values of $ a $ and $ k $.
\end{enumerate}

\subsection{\textbf{Darmour-Solodukhin wormhole}}
The Darmour-Solodukhin wormhole \cite{DS1} is a good candidate for cosmological observations, the ``black hole foils'', i.e. these wormholes are objects that mimic some aspects of black holes, while lacking some of their defining features, such as the event horizon. This is a modification of the well known Schwarzschild metric in order to make it horizonless. The stress energy tensor and the information for the source required to maintain this WH are discussed in \cite{DS2}, where the authors used the Schwinger-DeWitt expansion to derive an approximated stress-energy tensor of the quantized massive scalar, spinor and vector field for the DS WH. They found that for the scalar field there is a region in the parameter space for which the stress-energy tensor has the desired properties. The stress energy of the massive scalar field with a general curvature coupling $ \xi $ is:
\begin{equation}
T^{b}_{\,a}=\dfrac{1}{96\pi^{2}\mu^{2}m^{6}}\left(1-\frac{2}{x}-\lambda^2\right)^{-6} \sum_{k=0}^{7} \beta^{(k)b}_{a}\frac{1}{x^{k+8}}.
\end{equation}
Here $ \mu $ is the mass of the field with $ x=r/m $. The coefficients $ \beta^{(k)b}_{a} $ depend parametrically on $ \lambda $ and $ \xi $.
Considering this stress-energy tensor, they found that there is a region in the $(\lambda,\xi)-$plane in which the stress-energy tensor has the form required to support the wormhole. The exact form of the stress-energy tensor of quantized massive fields are complicated and can be found in the supplementary of the same paper.

\begin{figure}[ht]
    \centering
    \subfloat[Subfigure 1 list of figures text][]
        {
        \includegraphics[width=0.48\textwidth]{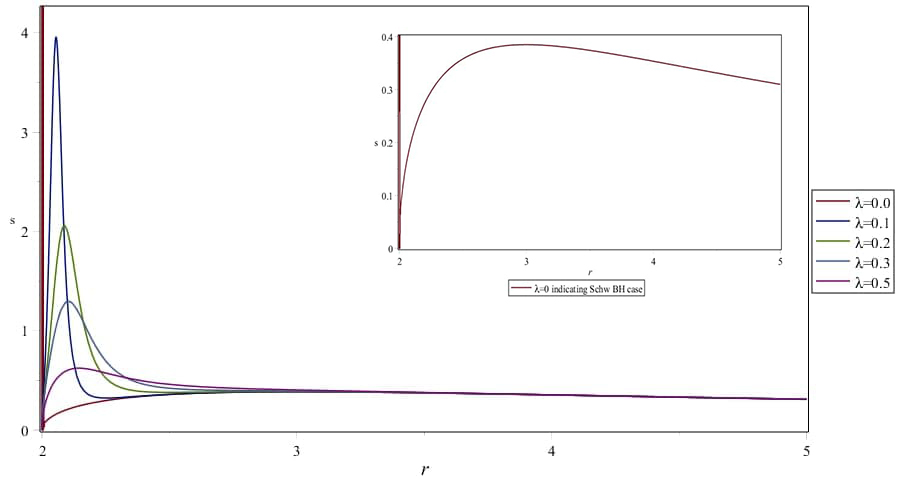}
        \label{fig:subfig3}
        }
    \subfloat[Subfigure 2 list of figures text][]
        {
        \includegraphics[width=0.48\textwidth]{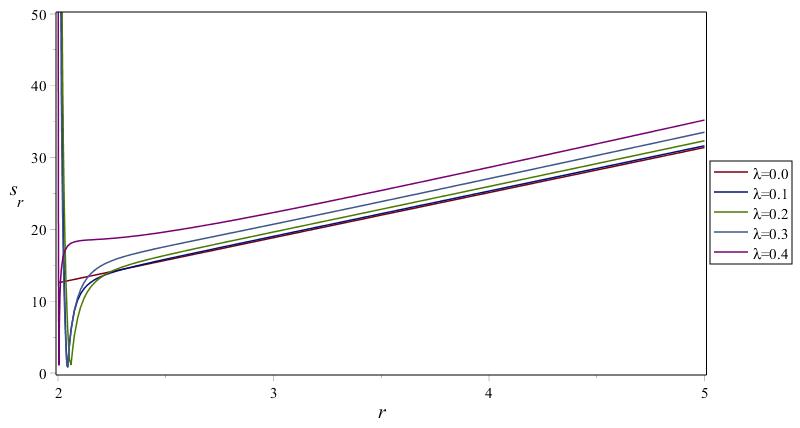}
        \label{fig:subfig4}
        }
    \caption{(a) Variation of GE density $ s $ of DS wormhole with $ r $ for various parameter strength $ \lambda $ using the Rudjord proposal. (b) Rate of variation of CET gravitational entropy  $ \partial_{r}s_{grav}\equiv s_{r} $ of DS wormhole with $ r $ for various parameter strength $ \lambda $ using the CET proposal.}
   \label{DSentrop}
\end{figure}

The Darmour-Solodukhin (DS) wormhole metric is
\begin{equation}
ds^2=-(f(r)+\lambda^2)dt^2+\dfrac{dr^2}{f(r)}+r^2(d\theta^2+\sin^2\theta d\phi^2),
\end{equation}
where $ f(r)=(1-\frac{2m}{r})$, and $ \lambda $ is a dimensionless parameter. Here, if $ \lambda=0 $, then the Darmour-Solodukhin wormhole metric reduces to the usual Schwarzschild black hole metric having an event horizon at $ r=2m $. But for $ \lambda \neq 0 $, however small, there is no event horizon. Instead this Lorentzian wormhole have a throat at $ r=2m $ connecting two isometric and asymtotically flat regions $ 2m\leq r \leq \infty $.
\begin{enumerate}
\item \textbf{Weyl scalar proposal:} In order to begin our analysis we first need to identify the spatial section of the Darmour-Solodukhin wormhole metric, which is given by the following expression:
$$ h_{ij}= \text{diag}\left(\dfrac{1}{(1-\frac{2m}{r})},r^2,r^2\sin^2\theta\right) .$$
We also need to compute the ratio $(P_{1}^2)$, which is given by the expression \eqref{DS1}:
\begin{align}\label{DS1}
\left.P_{1}^2\right. & =\dfrac{C_{abcd}C^{abcd}}{R_{abcd}R^{abcd}}\nonumber\\ & =\frac{1}{3} \left( 3\,{\lambda}^{4}{r}^{2}-19\,{\lambda}^{2}mr+9\,{
\lambda}^{2}{r}^{2}+24\,{m}^{2}-24\,mr+6\,{r}^{2} \right) ^{2} \times \nonumber\\
& \big(6\,{\lambda}^{8}{r}^{4}-48\,{\lambda}^{6}m{r}^{3}+24\,{\lambda}^{6}{r}^{4}
+177\,{\lambda}^{4}{m}^{2}{r}^{2}-172\,{\lambda}^{4}m{r}^{3}+42\,{\lambda}^{4}{r}^{4}-304\,{\lambda}^{2}{m}^{3}r+ \nonumber\\
& 448\,{\lambda}^{2}{m}^{2}{r}^{2}-220\,{\lambda}^{2}m{r}^{3}+36\,{\lambda}^{2}{r}^{4}+192\,{m}
^{4}-384\,{m}^{3}r+288\,{m}^{2}{r}^{2}-96\,m{r}^{3}+12\,{r}^{4}\big)^{-1}.
\end{align}
\begin{figure}[ht]
    \centering
\includegraphics[width=1.0\textwidth, height=0.32\textheight]{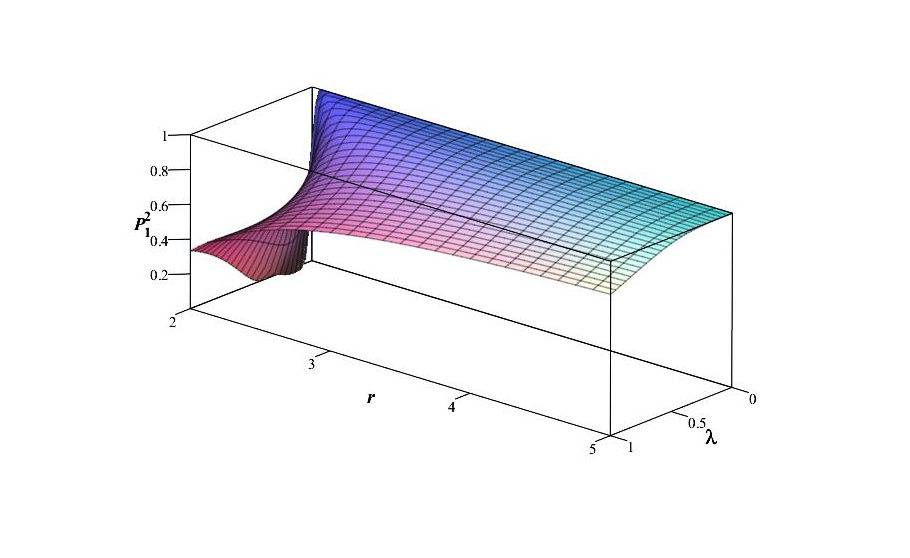}
\caption{Variation of the ratio of curvature scalars $P_{1}^2 $ as a function of $ r,\lambda $ for Darmour-Solodukhin WH.}
\label{DSW}
\end{figure}
The ratio of curvature scalars computed in \eqref{DS1} is depicted in Fig.\ref{DSW}. We find that as we approach the throat region at $ r=2 $, the contribution of the matter dominates over the Weyl, resulting in a dip of the value near that region, which is expected as the mass is located in that region. Using the proposal of Rudjord et al., we arrive at the expression of GE density in \eqref{DS2} which is depicted in Fig.\ref{DSentrop}(a). We note that the gravitational entropy density is peaking around the throat of the wormhole. Further, as the value of $ \lambda $ increases, the peak of the gravitational entropy density decreases, with the highest value in the case of the black hole itself.

\begin{align}\label{DS2}
\left.s=k_{s}\vert\nabla.\Psi\vert  \right. & = 3072k_{s}\,\sqrt {3} \Bigg| \frac {\sqrt {-r+2\,m}}{{r}^{3/2}} \Bigg( {\frac {{\lambda}^{12}{r}^{6}}{256}}-{\frac {79\,{r}^{5}{\lambda}^{10}}{1536} \left( m-{\frac {42\,r}{79}} \right) }+  \nonumber\\
 & \left( {\frac {329\,{m}^{2}{r}^{4}}{1024}}-{\frac {165\,m{r}^{5}}{512}}+{\frac {21\,{r}^{6}}{256}} \right) {\lambda}^{8}-\frac {583\,{r}^{3}{\lambda}^{6}}{512}\left( {m}^{3}-{\frac {5147\,{m}^{2}r}{3498}}+{\frac {423\,m{r}^{2}}{583}}-{\frac {70\,{r}^{3}}{583}} \right) +\nonumber\\
 & {\frac {41\,{r}^{2} \left(
-r/2+m \right) {\lambda}^{4}}{18} \left( {m}^{3}-{\frac {3837\,{m}^{2}
r}{2624}}+{\frac {3747\,m{r}^{2}}{5248}}-{\frac {153\,{r}^{3}}{1312}}
 \right) }- {\frac {19\,r \left( -r/2+m \right) ^{4}{\lambda}^{2}}{8}
 \left( m-{\frac {9\,r}{19}} \right) }+  \nonumber\\
 & \left( -r/2+m \right) ^{6}
 \Bigg)\Bigg( 6\,{\lambda}^{8}{r}^{4}+ \left( -48\,m{r}^{3}+24\,{r}
^{4} \right) {\lambda}^{6}+ \left( 177\,{m}^{2}{r}^{2}-172\,m{r}^{3}+
42\,{r}^{4} \right) {\lambda}^{4}- \nonumber\\
& 304\,r \left( -\frac{r}{2}+m \right) ^{2}
 \left( m-{\frac {9\,r}{19}} \right) {\lambda}^{2}+192\, \left( -\frac{r}{2}+m
 \right) ^{4} \Bigg) ^{-\frac{3}{2}} \Bigg|.
\end{align}
If we put $ \lambda\rightarrow 0 $ in \eqref{DS2}, then the entire expression for $ s $ reduces to that of the Schwarzschild black hole, i.e. $ s=\dfrac{2k_{s}}{r} \Big|\sqrt{1-\frac{2m}{r}}\Big|$, which matches with the result in \cite{entropy1}. This is depicted in Fig.\ref{DSentrop}(a), validating the expression we have obtained here. We note that the GE density for the Schwarzschild black hole has a maxima at $ r=\frac{3}{2}\times(2m)$ and goes to zero at the event horizon at $ r=2m $. We have taken $ m=1 $ for our plots.
\item \textbf{CET proposal:} Let us now consider the CET proposal of gravitational entropy. We have chosen the four vectors to be aligned with the principal null tetrads, and computed the Weyl curvature scalar $ \Psi_{2} $ as follows:
\begin{equation}
\Psi_{2}= -4\,{\frac {m}{ \left( -{\lambda}^{2}r+2\,m-r \right) ^{2}{r}^{3}}
 \left(  \left( 1/8\,{\lambda}^{4}+3/8\,{\lambda}^{2}+1/4 \right) {r}^
{2}+m \left( -{\frac {19\,{\lambda}^{2}}{24}}-1 \right) r+{m}^{2}
 \right) }.
\end{equation}
Now using the definition of gravitational energy density $ \rho_{grav} $ and the expression of $ \Psi_{2} $, we have obtained the following expression \eqref{DS3}:
\begin{equation}\label{DS3}
\rho_{grav}=\dfrac{\alpha}{4\pi}\vert\Psi_{2}\vert={\frac {\alpha}{24\pi} \left| {\frac {m \left( 3\,{\lambda}^{4}{r}
^{2}-19\,{\lambda}^{2}mr+9\,{\lambda}^{2}{r}^{2}+24\,{m}^{2}-24\,mr+6
\,{r}^{2} \right) }{ \left( {\lambda}^{2}r-2\,m+r \right) ^{2}{r}^{3}}
} \right| }.
\end{equation}
Similarly using the definition of gravitational temperature $ T_{grav} $ we obtain \eqref{DS4}:
\begin{equation}\label{DS4}
T_{grav}=\frac{1}{2\pi}\left\vert\dfrac{m\sqrt{r-2m}}{(r-2m+\lambda^2 r)r^{3/2}}\right\vert.
\end{equation}
Here we chose the equatorial plane for our calculations. Finally, we obtain the expression for the rate of variation of CET gravitational entropy along the radial coordinate given in \eqref{DS5} (depicted in Fig.\ref{DSentrop}(b)):
\begin{equation}\label{DS5}
\partial_{r}s_{grav}=\frac{\alpha \pi r}{3}\, \left| {\frac {3\,{\lambda}^{4}{r}^{2}-19\,{\lambda}^{2
}mr+9\,{\lambda}^{2}{r}^{2}+24\,{m}^{2}-24\,mr+6\,{r}^{2}}{(r-2
\,m) \left( {\lambda}^{2}r-2\,m+r \right) }} \right|.
\end{equation}
The variation of the local piecewise gravitational entropy of the DS wormhole is obtained as:
\begin{align}
& s_{grav}(r)=\dfrac{4\pi}{\left( {\lambda}^{2}+1 \right) ^{2}}\times{\it signum}
 \left( {\frac { \left( 3\,{\lambda}^{4}+9\,{\lambda}^{2}+6 \right) {r
}^{2}+ \left( -19\,{\lambda}^{2}-24 \right) mr+24\,{m}^{2}}{ \left( -r
+2\,m \right)  \left( -{\lambda}^{2}r+2\,m-r \right) }} \right) \nonumber \\
& \bigg( \frac{{m}^{2}}{6}\ln  \left( {\lambda}^{2}r-2m+r
 \right) + \left( {m}^{2} \left( {\lambda}^{2}+1 \right)  \left( {
\lambda}^{2}-\frac{1}{6} \right) \ln  \left( r-2m \right) +\frac{r}{2} \left(
 \left( m+\frac{r}{4} \right) {\lambda}^{4}+ \left( \frac{5}{6}m+\frac{3}{4}r \right) {
\lambda}^{2}+\frac{r}{2} \right)  \right) \nonumber \\
& \left( {\lambda}^{2}+1 \right)
 \bigg)\bigg\vert_{2m+\epsilon}^{r},
\end{align}
where $\epsilon>0$ is a very small quantity as the above expression is not valid for $ r=2m $.
\end{enumerate}
We find that in this case, the rate of variation of gravitational entropy along the radial direction, $\partial_{r}s_{grav}$, decreases as the throat region is approached. Over here, as the value of $ \lambda $ increases, the magnitude goes towards a higher value. We also observe that the rate of variation of CET GE goes towards zero near the throat region for nonzero $ \lambda $.

\subsection{\textbf{Exponential metric wormhole}}
Next we will analyze the exponential metric wormhole. The exponential or Papapetrou metric represents the counterpart to the Schwarzschild black hole with antiscalar background fields, and can be considered as a special case of the Fisher–(Newman–Janis–Winicour)–Wyman–Ellis–Bronnikov solutions for a massless scalar field coupled to gravity. It is related to the Morris–Thorne traversable wormhole. A detailed description of the antiscalar source field can be found in \cite{exp2,exp4}, where the authors analysed the origin of the metric within a wide class of scalar and antiscalar solutions of the Einstein equations parameterized by scalar charge. The exponential metric wormhole in isotropic coordinates is given by :
\begin{equation}
ds^2=-e^{\frac{-2m}{r}}dt^2+e^{\frac{2m}{r}}[dr^2+r^2(d\theta^2+\sin^2\theta d\phi^2)],
\end{equation}
with the Einstein tensor $ G^{a}_{\,b}=\dfrac{m^{2}e^{\dfrac{-2m}{r}}}{r^4}\text{diag}\left\lbrace 1,-1,1,1 \right\rbrace^{a}_{\,b}  $, and the antiscalar source field described by: $ T^{SF}_{ab}=\frac{1}{4\pi}\left(   \phi_{a}\phi_{b}-\frac{1}{2}g_{ab}\phi_{c}\phi^{c}\right) $. The energy momentum tensor is quadratic in $ \phi_{a}=\nabla_{a}\phi=\phi_{,a} $, with the field equation $ G_{ab}=-8\pi T^{SF}_{ab} $. The Lagrangian for the antiscalar field is given by $ L=\frac{1}{16\pi}\left(R+2\phi_{a}\phi^{a}\right) $. Recently, Boonserm et.al. \cite{exp1} demonstrated that this metric represents a traversable wormhole. Here $ m $ is the mass of the wormhole. The throat of the WH is located at $ r=m $. This WH does not have any horizon because $ g_{tt} $ is nonzero for all non negative values of $ r\in(0,+\infty) $. The region $ r<m $ represents the other universe, and the curvature invariants are nonzero at the throat \cite{WHi} while they go to zero asymptotically as $ r $ goes to infinity (as it should), in order to connect two asymptotically flat regions. In \cite{exp3} the authors have studied the geodesics using the Jacobi metric approach.
\begin{enumerate}
\item \textbf{Weyl scalar proposal:} For the exponential metric wormhole we find that the spatial section is the following:
\begin{equation}
h_{ij}= e^{\frac{2m}{r}} \text{diag}(1,r^2,r^2 \sin^2\theta),
\end{equation}
where the determinant of $ h_{ij} $ is given by: $ h= e^{\frac{6m}{r}}r^4 \sin^2\theta$. The ratio of the curvature scalars is given in \eqref{expw}:
\begin{equation}\label{expw}
P_1^2=\frac{4}{3}\dfrac{(2m-3r)^2}{(7m^2-16mr+12r^2)}.
\end{equation}
From this expression it is clear this ratio becomes zero at $ r=\frac{2m}{3} $. For the sake of clarity, the variation of $ P_{1}^2 $ is shown in FIG.\ref{EXPP} for different values of mass parameter $ m $. From the figure it is clear that $ P_{1}^2 $ becomes zero at specific points as mentioned above, and there is also a dip in the values near the throat of the wormhole where the mass is located, implying a more dominant contribution from the Riemann tensor than the Weyl component.
\begin{figure}[ht]
    \centering
\includegraphics[width=0.65\textwidth]{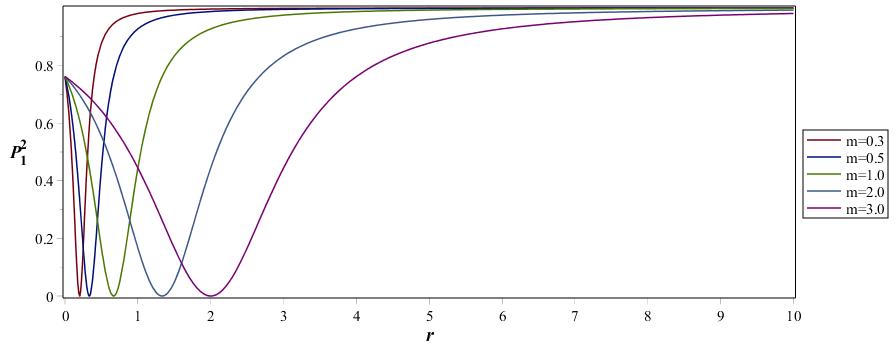}
\caption{ Variation of the ratio of curvature scalars $P_{1}^2 $ as a function of $ r $ for different $ m $ in the case of exponential metric wormhole.}
\label{EXPP}
\end{figure}
In this case we will use the following definition of gravitational entropy density:
\begin{equation}\label{e2}
s=k_{s}\vert\nabla.\Psi \vert= k_{s}\left\vert\frac{1}{\sqrt{h}}\dfrac{\partial}{\partial r}(\sqrt{h}\dfrac{P_{1}}{\sqrt{h_{rr}}}) \right\vert.
\end{equation}
\begin{figure}[ht]
    \centering
    \subfloat[Subfigure 1 list of figures text][]
        {
        \includegraphics[width=0.46\textwidth]{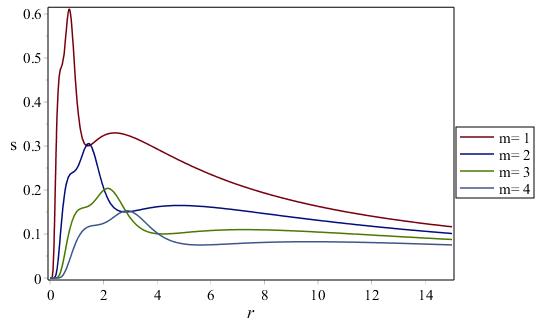}
        \label{fig:subfig3a}
        }
    \subfloat[Subfigure 2 list of figures text][]
        {
        \includegraphics[width=0.46\textwidth]{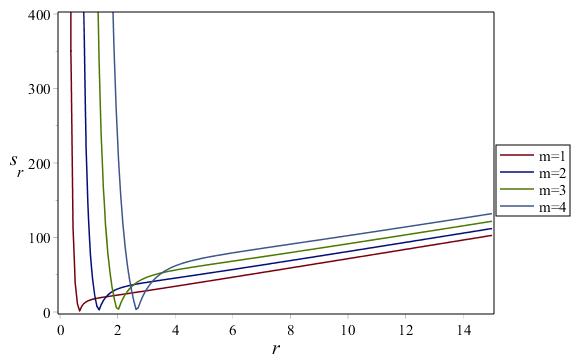}
        \label{fig:subfig4a}
        }
    \caption{(a)Variation of GE density $ s $ of exponential metric wormhole with $ r $ for various throat lengths $ r=m $ using Rudjord proposal. (b) Rate of variation of gravitational entropy along the radial direction $ \partial_{r}s_{grav}\equiv s_{r} $ of exponential metric wormhole with $ r $ for various throat lengths $ r=m $ using CET proposal.}
   \label{expentrop}
\end{figure}
The above definition of gravitational entropy density yields the result in \eqref{exps1}.
\begin{equation}\label{exps1}
   s = \frac{2k_{s}}{3}  \frac {\sqrt {3}}{{r}^{2} \left( 7\,{m}^{2}-16\,mr+12\,{r}^{2}
 \right) ^{2}} \left| {{\frac {(56\,{m}^{5}-352\,{m}^{4}r+912\,{m}^{3}{r}^{2}-
1197\,{m}^{2}{r}^{3}+792\,m{r}^{4}-216\,{r}^{5})}{\sqrt {{
\frac { \left( 2\,m-3\,r \right) ^{2}}{7\,{m}^{2}-16\,mr+12\,{r}^{2}}}
}}}} \right|  \left( {{\rm e}^{{\frac {m}{r}}}} \right) ^{-1}.
\end{equation}
In Fig.\ref{expentrop}(a) we have shown the variation of gravitational entropy density using  \eqref{exps1} with different wormhole throat values. We find that the entropy density increases and becomes maximum near the throat and decays to zero as we move away from the central throat region. As the mass increases, so does the throat radius and the height of the maximum of the gravitational entropy density decreases with it, which is expected. Also the entropy density goes to zero on the other side of the throat as it approaches $ r=0. $
\item \textbf{CET proposal:} Next we consider the CET gravitational entropy proposal. At first the gravitational epoch function $ w $ is calculated:
\begin{equation}
w=T_{tttt}u^{t}u^{t}u^{t}u^{t}=\dfrac{m^2 e^{\frac{-4m}{r}}(2m-3r)^2}{6r^8}.
\end{equation}
Once again all the Weyl scalars are calculated independently and $ \Psi_{2} $ turns out to be the only nonzero component which is given below:
\begin{equation}
\Psi_{2}=\dfrac{m e^{\frac{-2m}{r}}(2m-3r)}{3r^4}.
\end{equation}
Here also the identity $ \vert \Psi_{2} \vert=\sqrt{\dfrac{2w}{3}} $ is satisfied as this is a Petrov type D spacetime. The gravitational energy density $\rho_{grav}  $ is given by
\begin{equation}
\rho_{grav}=\frac{\alpha}{4\pi}\vert\Psi_{2}\vert=\frac{\alpha}{12\pi} \left\vert\dfrac{m e^{\frac{-2m}{r}}(2m-3r)}{r^4}\right\vert.
\end{equation}
As $ \dot u_{r}=\dfrac{m}{r^2}, \Theta=0, \sigma_{ab}=0  $ and $ \omega_{ab}=0 $, we can calculate the gravitational temperature as
\begin{equation}
T_{grav}={\dfrac{1}{2\pi}}\left\vert \dfrac{m}{r^2 e^{\frac{m}{r}}}\right\vert.
\end{equation}
Once again we consider the equatorial plane, and compute the rate of variation of GE along the radial direction to get the following expression:
\begin{equation}\label{sgravcet}
\partial_{r}s_{grav}=\dfrac{\rho_{grav}v}{T_{grav}}=\frac{2\alpha\pi}{3} \left\vert e^{\frac{2m}{r}}(2m-3r)\right\vert.
\end{equation}
Finally, we obtain the piecewise local gravitational entropy of the exponential metric wormhole as the following:\\
\begin{equation}
s_{grav}(r)=\int_{0}^{r}\dfrac{\rho_{grav}v}{T_{grav}}dr=\frac {2\alpha\pi}{3}
\begin{cases}
\left\vert-\frac{3}{2}\,{{\rm e}^{{\frac {2m}{r}}}}{r}^{2}-{{\rm e}^{{\frac {2m}{r}}}}rm-2\,{\it Ei} \left( 1,-{\frac {2m}{r}} \right) {m}^{2}\right\vert; \quad r< \frac{2m}{3},\\
\it \text{undefined}; \quad r=\frac{2m}{3},\\
\left\vert\frac{3}{2}\,{{\rm e}^{{\frac {2m}{r}}}}{r}^{2}+{{\rm e}^{{\frac {2m}{r}}}}rm+2\,{\it Ei} \left( 1,-{\frac {2m}{r}} \right) {m}^{2}\right\vert; \quad \frac{2m}{3}<r.
\end{cases}
\end{equation}
where $ Ei(a,z) $ are exponential integrals. In Fig.\ref{expentrop}(b), the rate of variation of gravitational entropy along the radial direction is plotted using \eqref{sgravcet} for different wormhole throat values. The zeroes are the points where the Weyl curvature becomes zero (as also in \eqref{expw}), i.e. at $ r=\frac{2m}{3} $. Therefore, as $ m $ increases, the throat radius increases, and the zero of the rate of radial variation of GE in the CET convention shifts toward positive infinity. As the distance increases from the central throat, i.e., at $r=m$ to the other side, it rises sharply from zero. This is also evident from equation \eqref{sgravcet}. A similar behaviour can also be found for $ s_{grav} $: it decreases monotonically as we approach the throat, crosses the throat smoothly, and then sharply increases on the other side.
\end{enumerate}

\subsection{\textbf{Traversable NUT wormhole}}
The solution of the Einstein-Maxwell system of equations found by Brill \cite{brill} in $1964$ is given by
\begin{equation}\label{trav_NUT_wmhl}
ds^2=-f(dt-2n(cos\theta +C)d\phi)^2+f^{-1}dr^2+(r^2+n^2)(d\theta^2+sin^2\theta d\phi^2).
\end{equation}
The Brill solution with no horizon, which connects two asymptotically locally flat regions, is the wormhole of our interest \cite{WH1}. The unknown parameters are as follows:
$$ f= \dfrac{(r-m)^2+b^2}{r^2+n^2}, \quad \textrm{and} \quad b^2=q^2+p^2-m^2-n^2 =e_{2}-m^2-n^2, $$
where $ n $ is the NUT parameter, $ m $ is the mass parameter, $ q $ and $ p $ are the electric and magnetic charges respectively. For the sake of simplicity we will combine these two charges and call them as $ e_{2}= q^2+p^2$.
For $b^2<0 $ it has two horizons, just as the Reissner-Nordstr\"{o}m (RN) solution. For $ b^2=0 $, it has, just as the extreme RN solution, a double horizon. However, for $b^2>0$, contrary to the RN solution, it is not singular, but has the (Lorentzian) wormhole topology, the coordinate $ r $ varying along the whole real axis, with two asymptotic regions $r = \pm \infty$. As $r (> 0)$ decreases, the $r =$ constant $2$-spheres shrink until a minimal sphere of area $4 \pi n^2$ (the wormhole neck) is reached for $r=0$, and then expand as $r (< 0)$ continues to decrease.
The determinant of the metric \eqref{trav_NUT_wmhl} for the above mentioned traversable NUT wormhole is given by
$ g=- \left( \sin \left( \theta \right)  \right) ^{2} \left( {n}^{4}+2\,{n}^{2}{r}^{2}+{r}^{4} \right). $\\
Subsequently the Weyl scalar is given by the expression:
\begin{eqnarray}
W= -\frac{48}{\left( {n}^{2}+{r}^{2}
 \right) ^{6}} \left( -{n}^{4}+ \left( m-3\,r \right) {n}^{3}+ \left( -
3\,mr+3\,{r}^{2}+e_{{2}} \right) {n}^{2}+ \left( -3\,m{r}^{2}+{r}^{3}+
2\,re_{{2}} \right) n+{r}^{2} \left( mr-e_{{2}} \right)  \right) \nonumber \\
\left( {n}^{4}+ \left( m-3\,r \right) {n}^{3}+ \left( 3\,mr-3\,{r}^{2
}-e_{{2}} \right) {n}^{2}+ \left( -3\,m{r}^{2}+{r}^{3}+2\,re_{{2}}
 \right) n-m{r}^{3}+e_{{2}}{r}^{2} \right).
\end{eqnarray}
\\
\begin{enumerate}
\item \textbf{Weyl scalar proposal:} To evaluate the gravitational entropy density of this spacetime we need the Kretschmann curvature scalar for this metric, which is the following:
\begin{equation}
 K=-\frac{8}{\left( {n}^{2}+{r}^{2} \right) ^{6}} \Sigma .
\end{equation}
The square root of the ratio of the Weyl scalar and the Kretschmann curvature scalar gives us the magnitude of the vector $\boldsymbol{\Psi}$, since $P_{1}^{2}=\dfrac{W}{K} $. Thus $ P_{1} $ is obtained in the form
\begin{align}
\left. P_{1}\right.&=\sqrt{\left( -{n}^{4}+ \left( m-3\,r \right) {n}^{3
}+ \left( -3\,mr+3\,{r}^{2}+e_{{2}} \right) {n}^{2}+ \left( -3\,m{r}^{
2}+{r}^{3}+2\,re_{{2}} \right) n+{r}^{2} \left( mr-e_{{2}} \right)
 \right) } \times \nonumber\\
  & \sqrt{6}\sqrt{\left( {n}^{4}+ \left( m-3\,r \right) {n}^{3}+ \left( 3\,mr-
3\,{r}^{2}-e_{{2}} \right) {n}^{2}+ \left( -3\,m{r}^{2}+{r}^{3}+2\,re_
{{2}} \right) n-m{r}^{3}+e_{{2}}{r}^{2} \right)} \times \Sigma^{-\frac{1}{2}} ,
\end{align}
where $ \Sigma $ is given by the following expression:
\begin{align}
\left. \Sigma\right.&=\big[  6\,{m}^{2}({n}^{6}-15\,{n}^{4}{r}^{2}+15\,{n}
^{2}{r}^{4}-{r}^{6})-24\,m(3{n}^{6}r-10{n}^{4}{r}^{3}+3
{n}^{2}{r}^{5})-6\,{n}^{2}({n}^{6}-15\,{n}^{4}{r}^{2}+15\,{n}^{2}{r}^{4}-{r}^{6}) \nonumber\\
&+e_{{2}}(60\,m{n}^{4}r-120\,m{n}^{2}{r}^{3}+12\,m{r}^
{5}+12\,{n}^{6}-120\,{n}^{4}{r}^{2}+60\,{n}^{2}{r
}^{4}-7\,{n}^{4}{e_{{2}}}+34\,{n}^{2}{r}^{2}{e_{{2}}}-7
\,{r}^{4}{e_{{2}}} )\big].
\end{align}
\\
\begin{figure}[ht]
    \centering
    \subfloat[Subfigure 1 list of figures text][]
        {
        \includegraphics[width=0.48\textwidth, height=0.16\textheight]{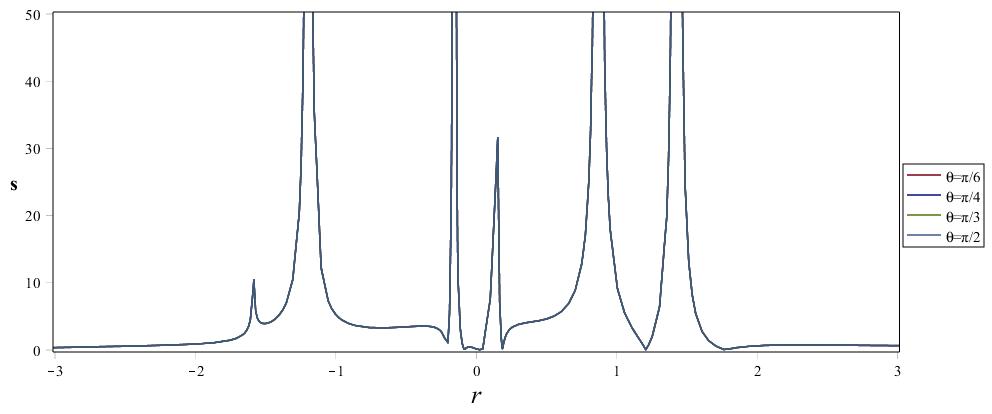}
        \label{fig:subfig3b}
        }
    \subfloat[Subfigure 2 list of figures text][]
        {
        \includegraphics[width=0.48\textwidth]{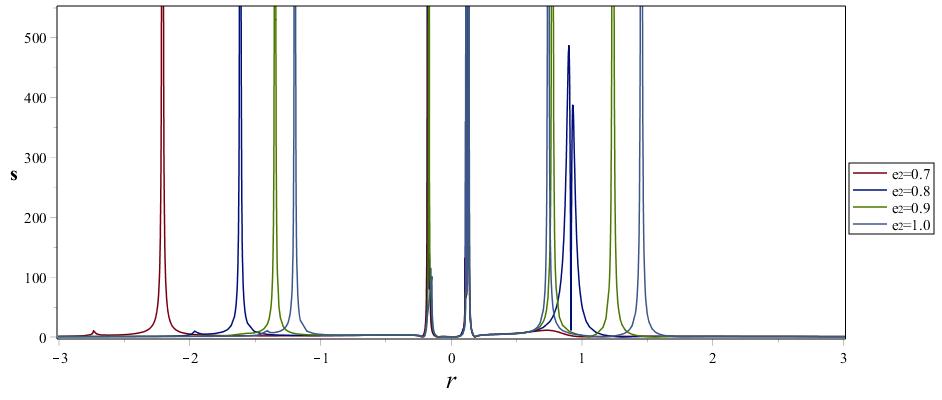}
        \label{fig:subfig4b}
        }
    \caption{Variation of GE density $ s $ of traversable NUT wormhole with $ r $ for different parameters. Here we have only considered the radial contribution as in \eqref{sd1}. In this case the expression of GE density have been calculated using $ P_{1} $.}
   \label{NUT1r}
\end{figure}



\begin{figure}[ht]
    \centering
    \subfloat[Subfigure 1 list of figures text][]
        {
        \includegraphics[width=0.48\textwidth]{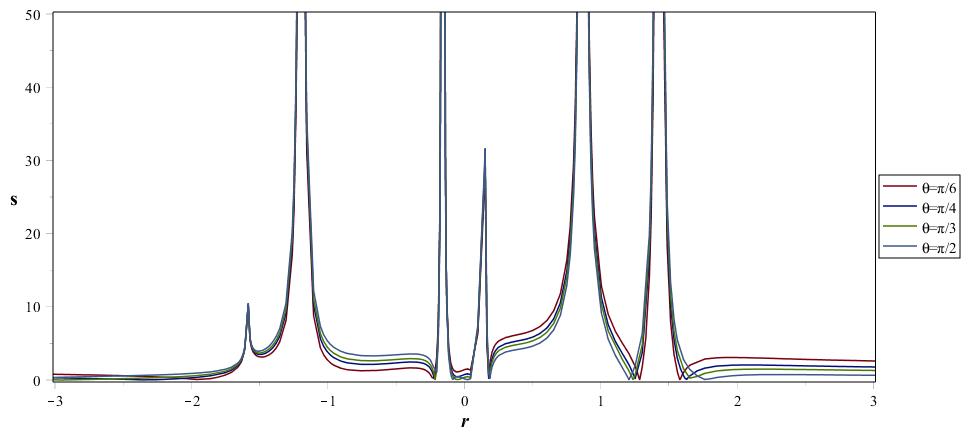}
        \label{fig:subfig3c}
        }
    \subfloat[Subfigure 2 list of figures text][]
        {
        \includegraphics[width=0.48\textwidth, height=0.165\textheight]{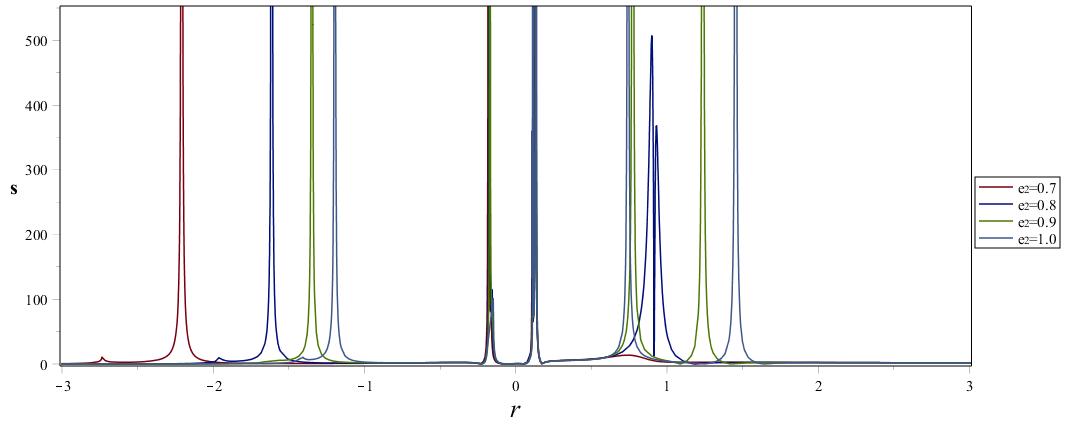}
        \label{fig:subfig4c}
        }
    \caption{Variation of GE density $ s $ of traversable NUT wormhole with $ r $ for different parameters. Here we have considered both the radial and angular contribution as in \eqref{sd2}. Here also the expression of GE density is being calculated using $ P_{1} $.}
   \label{NUT1rt}
\end{figure}





In FIG. \ref{NUT1r} and FIG. \ref{NUT1rt}, the gravitational entropy density for traversable NUT wormhole have been illustrated using the definition involving $ P_{1} $. To generate FIG. \ref{NUT1r}, we have considered only the radial component in the definition of gravitational entropy density, i.e., using \eqref{sd1}, whereas in FIG. \ref{NUT1rt} both the radial and angular contributions have been considered using \eqref{sd2}, as the metric has a nonzero $ g_{t\phi} $ component.

In FIG. \ref{NUT1r}(a), the GE density have been examined by varying the angular orientation $ \theta $. In this case we have taken $ m=0.2,\, n=0.3 $ and $ e_{2}=1 $.
In FIG. \ref{NUT1r}(b) the variation of the GE density with the parameter $ e_{2} $ (charge squared) is shown for the same values of the parameters, i.e. $ m=0.3,\, n=0.3 $ and with $ \theta=\frac{\pi}{4} $.

FIG. \ref{NUT1rt}(a) is drawn by assuming $ m=0.2,\, n=0.3 $ and $ e_{2}=1 $ to show the variation of the gravitational entropy density with angular orientation $ \theta $.
In the second case, i.e. in FIG. \ref{NUT1rt}(b), the effect of charge has been shown, where we have taken $ m=0.3,\, n=0.3 $ and $ \theta=\frac{\pi}{4} $ as our fixed parameters.
From both FIG. \ref{NUT1r} and FIG. \ref{NUT1rt}, which are drawn by considering the expression for $ P_{1} $, it is evident that there are too many discontinuities. This means that $ P_{1} $ does not serve as a good measure of gravitational entropy density in this case. Therefore we considered the alternative proposal with $ P_{2}=C_{abcd}C^{abcd} $ to calculate the GE. Here we will not quote the exact expressions for the GE density since these are extremely complicated and lengthy, but an idea regarding the nature of these expressions can be gathered if one refers to our work on accelerating black holes \cite{GC}.


\begin{figure}[ht]
    \centering
    \subfloat[]{\begin{minipage}[t][9cm][b]{.5\textwidth}
\centering
\includegraphics[width=1.0\textwidth, height=0.16\textheight]{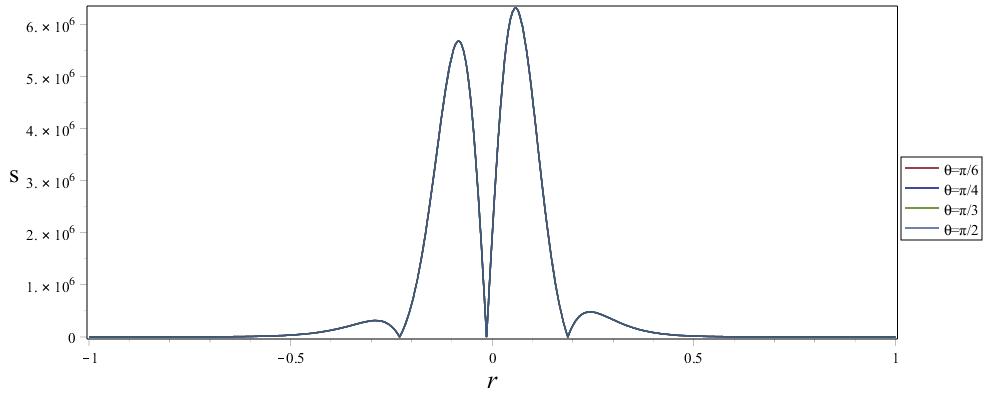}

\includegraphics[width=1.0\textwidth, height=0.16\textheight]{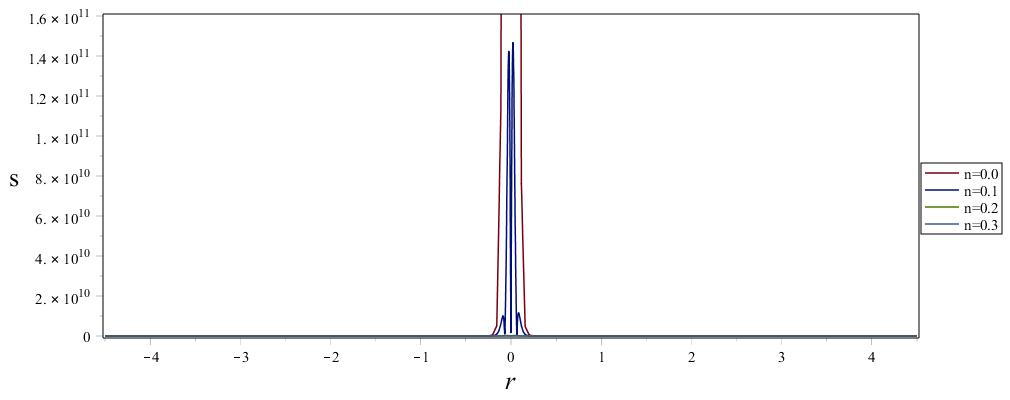}
\end{minipage}}%
       \subfloat[]{\begin{minipage}[t][9cm][b]{.5\textwidth}
\centering
\includegraphics[width=1.0\textwidth, height=0.16\textheight]{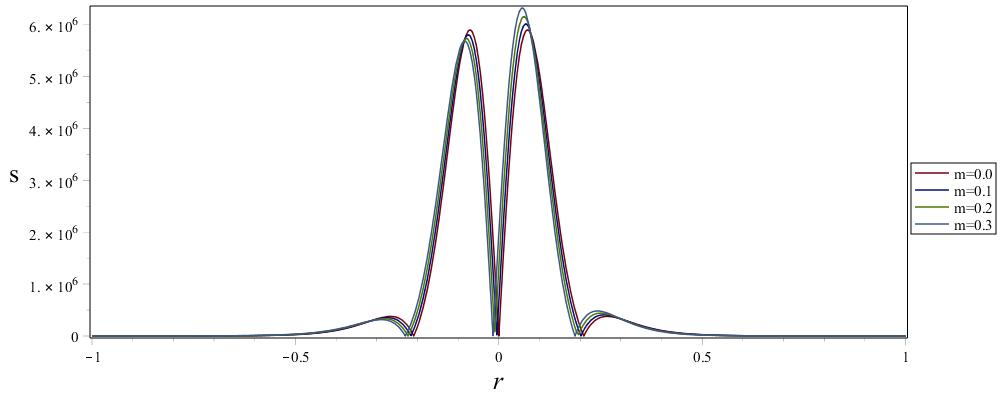}

\includegraphics[width=1.0\textwidth, height=0.16\textheight]{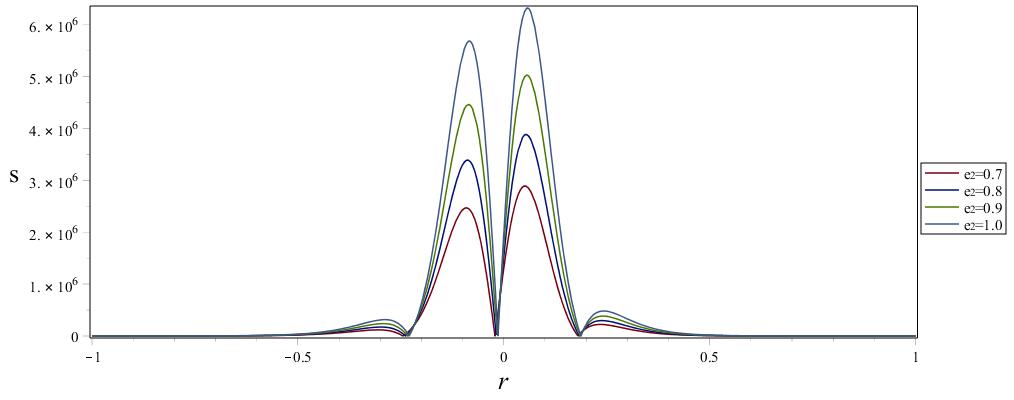}
\end{minipage}}%

    \caption{Variation of GE density $ s $ of traversable NUT wormhole with $ r $ for different parameters. Here we have only considered the radial contribution as in \eqref{sd1}. The definition of GE density is calculated using $ P_{2} $.}
   \label{NUT2r}
\end{figure}
\begin{figure}[ht]
    \centering
    \subfloat[]{\begin{minipage}[t][9cm][b]{.5\textwidth}
\centering
\includegraphics[width=1.0\textwidth, height=0.16\textheight]{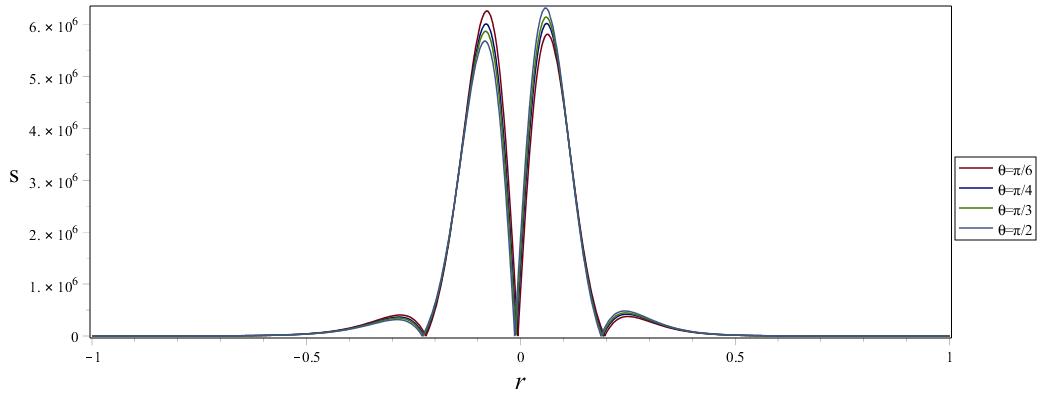}

\includegraphics[width=1.0\textwidth, height=0.16\textheight]{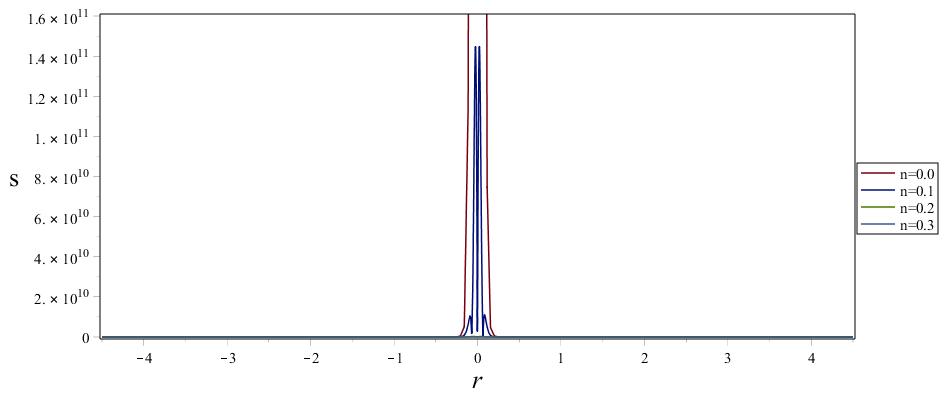}
\end{minipage}}%
       \subfloat[]{\begin{minipage}[t][9cm][b]{.5\textwidth}
\centering
\includegraphics[width=1.0\textwidth, height=0.16\textheight]{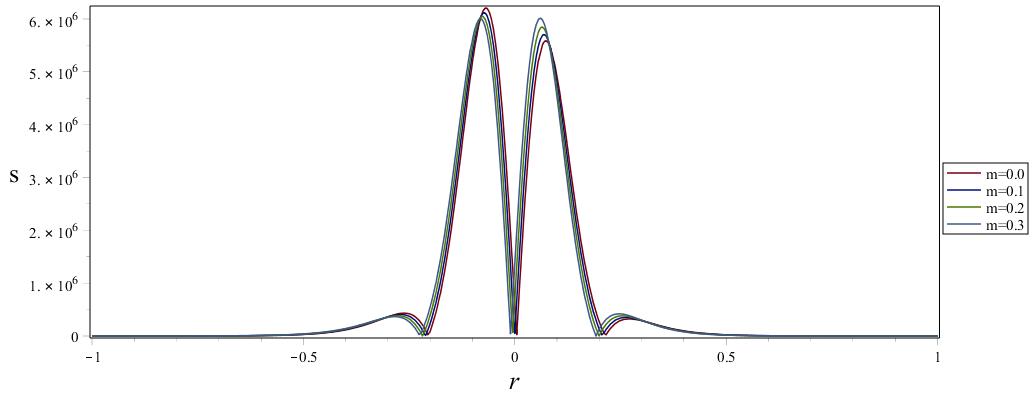}

\includegraphics[width=1.0\textwidth, height=0.16\textheight]{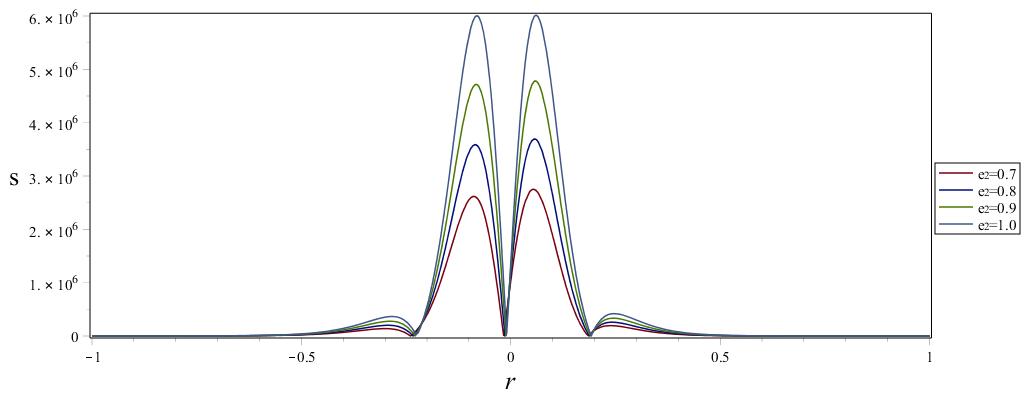}
\end{minipage}}%

    \caption{Variation of GE density $ s $ of traversable NUT wormhole with $ r $ for different parameters. Here we have considered both the radial and angular contribution as in \eqref{sd2}. The definition of GE density have been calculated using $ P_{2} $.}
   \label{NUT2rt}
\end{figure}

FIG. \ref{NUT2r} and FIG. \ref{NUT2rt} shows the investigation of the NUT wormhole using the expression \eqref{p2} for $ P_{2} $ in the definition of gravitational entropy density mentioned in \eqref{sd1} and \eqref{sd2} respectively.

FIG. \ref{NUT2r}(a) shows us the variation of GE density with angular orientation $ \theta $ (for fixed parameters $ m=0.2,\, n=0.3 $ and $ e_{2}=1 $) and the variation with NUT parameter while having fixed parameters as $ m=0.3,\, e_{2}=1 $ and $ \theta=\frac{\pi}{4} $. Similarly in FIG. \ref{NUT2r}(b), the top figure shows the variation of GE density with mass for $ n=0.3,\, e_{2}=1 $ and $ \theta=\frac{\pi}{4} $. The bottom figure of FIG. \ref{NUT2r}(b) gives us the effect of charge on the GE density for a traversable NUT wormhole with $ m=0.3,\, n=0.3 $ and $ \theta=\frac{\pi}{4} $. In FIG. \ref{NUT2r}, all the calculations are done by considering only the radial contribution in the definition of the GE density. In FIG. \ref{NUT2rt} all the figures are drawn by considering both the radial and angular components, which indicates small differences between the two cases. Parameters $ m=0.3,\, n=0.3 $ and $ e_{2}=1 $ are fixed in the top figure of FIG. \ref{NUT2rt}(a), which shows the variation with $ \theta $. In the bottom figure of the same column, the variation with NUT parameter is shown with parameters fixed as $ m=0.3,\, e_{2}=1 $ and $ \theta=\frac{\pi}{4} $. Finally in the column of FIG. \ref{NUT2rt}(b), the variation with mass and charge are shown for $ n=0.3,\, e_{2}=1 $, $ \, \theta=\frac{\pi}{4} $ and $ m=0.3,\, n=0.3 $, $\, \theta=\frac{\pi}{4} $ as the fixed parameters for these cases respectively.
\\
In all the above cases we note that the GE density rises near the throat region and vanishes in the asymptotic limit away from the wormhole. As the GE density function is a polynomial in $ r $ (in fact the curvature scalars themselves are such), for our chosen parameters it has at least two real roots. That is why we encounter two zeroes in the GE density function, meaning that the GE density is localized in two adjacent regions and the location of these regions depend entirely on the parameters of the wormhole, although maximum entropy occurs near the throat region only. From the above analysis we can clearly state that the Weyl scalar proposal involving $ P_{1} $ is not suitable for the analysis of this wormhole, but if we take $ P_{2} $ into consideration, then we can obtain a viable measure of GE density.
\item \textbf{CET proposal:} This spacetime is not strictly of Petrov type D, and hence the CET proposal cannot provide us a unique form of $ s_{grav} $ for this wormhole, as mentioned in the previous section when we provided an overview of the CET proposal.
\end{enumerate}


{$\bullet$} \textbf{NOTE:} In addition to the above mentioned traversable wormholes we have also analyzed and discussed two more types of traversable wormholes, namely, the AdS wormhole, and the very recently proposed wormhole ansatz by Juan Maldacena. Details of these two supplementary cases are presented in the Appendix section of this paper.

\section{Tolman law in wormhole spacetimes}
The Tolman law describes the spatial dependence of locally measured temperature distribution in a gravitational field \cite{SantiagoVisser}.
Of the four WHs that we considered till now, three are static spherically symmetric cases and they are in equilibrium. To expand our study on these systems we apply the famous Tolman law given by Richard Tolman in 1930 \cite{tolman}, which says that thermal equilibrium can exist within a temperature gradient provided that a gravitational field is present. Tolman assumed a sphere of fluid with the line element in the form
\begin{equation}\label{template}
ds^2=-e^{\mu}(dr^2+r^2 d\theta^2+r^2 sin^2\theta d\phi^2)+e^{\nu}dt^2 ,
\end{equation}
where $\mu$ and $\nu$ are functions of $r$. This represents the geometry of perfect fluid having spherical symmetry in full generality, and also the system is static, similar to the cases that we considered. The Tolman law, i.e., the relation between gravitational potential and equilibrium temperature measured by a local observer in proper coordinates, $ T_{0} $, is expressed as follows:
\begin{equation}\label{tol1}
\dfrac{d\, ln T_{0}}{dr}=-\frac{1}{2}\dfrac{d\nu}{dr}.
\end{equation}
In other words, the Tolman law shows that thermodynamic equilibrium in general relativistic spacetimes requires a temperature gradient, and a 4-acceleration to stop free fall. This law provides us with a way to compute temperature variation for a given stationary spacetime along a 4-velocity parallel to a timelike Killing vector field. The extension of this result was presented by R. C. Tolman and P. Ehrenfest in \cite{toleh}. They argued that
\begin{equation}\label{tol2}
T_{0}\sqrt{g_{00}}=\tilde{T}=\textsf{const.}
\end{equation}
i.e., the proper temperature measured by a local observer in thermal equilibrium depends on the position as given in the relation \eqref{tol2}, so that the product remains constant throughout the system. The quantity $ \tilde{T} $ remains constant for that system, and is called the ``Tolman temperature''. Let us now briefly examine the static WHs which we considered so far.
\\~~~~\\
\textbf{Ellis WH:}
Comparing the metric \eqref{ellism} of the Ellis WH with the general spherically symmetric spacetime metric \eqref{template}, and using the relations \eqref{tol1}, \eqref{tol2} we get $ T_{0} $, which is given below:
\begin{equation}
\dfrac{d\, ln T_{0}}{dr}=0 \Rightarrow T_{0}=\textsf{const.}=\tilde{T}.
\end{equation}
We find that the there is no gradient in local temperature. Here the 4-velocity of the static frame is a geodesic field which gives a constant temperature from Tolman's law. Hence the system is in Tolman temperature throughout.
\textbf{Darmour-Solodukhin WH:} Next we consider the DS wormhole. The temperature measured by the local observer is given by the expression \eqref{DST0}:
\begin{equation}\label{DST0}
\frac{1}{T_{0}}\dfrac{dT_{0}}{dr}=-\dfrac{m}{r^2(f(r)+\lambda^2)}\Rightarrow T_{0}=\dfrac{\mathbf{B}}{\sqrt{f(r)+\lambda^2}}.
\end{equation}
Here $\mathbf{B}$ is an integration constant. We can also validate the result \eqref{tol2} and express the local temperature in terms of Tolman temperature as given in \eqref{DST01}:
\begin{equation}\label{DST01}
T_{0}\sqrt{g_{00}}=\mathbf{B}=\tilde{T}\Rightarrow T_{0}=\dfrac{\tilde{T}}{\sqrt{f(r)+\lambda^2}}.
\end{equation}
Here we have the option to find the $ \tilde{T} $. As this metric reduces to the Schwarzschild metric for $ \lambda=0 $, the limit $ T_{0}(\infty)=\dfrac{\tilde{T}}{\sqrt{1+\lambda^2}} $, reduces to Hawking-Bekenstein temperature for $ \lambda=0 $. Therefore, $ \tilde{T}=T_{BH} $, and consequently $ T_{0}=\dfrac{T_{BH}}{\sqrt{f(r)+\lambda^2}} $ gives us the local temperature in terms of the Hawking-Bekenstein temperature $ (T_{BH}) $. Similarly at the throat, the temperature measured by the local observer is $ T_{0}(2m)=\frac{T_{BH}}{\lambda} $, which blows up at the event horizon for the Schwarzschild BH.
\\
\textbf{Exponential metric WH:} Finally we will discuss the exponential metric case. Like the previous cases we calculate the temperature measured by the local observer and find that
\begin{equation}
\frac{1}{T_{0}}\dfrac{dT_{0}}{dr}=-\frac{m}{r^2}\Rightarrow T_{0}=\mathbf{C} e^{\frac{m}{r}}.
\end{equation}
Here $ \mathbf{C} $ is the integration constant and it is indeed the Tolman temperature. In \eqref{EXPT01} we have expressed the local temperature in terms of the Tolman temperature:
\begin{equation}\label{EXPT01}
T_{0}\sqrt{g_{00}}=\mathbf{C}=\tilde{T}\Rightarrow
T_{0}=\tilde{T}e^{\frac{m}{r}}.
\end{equation}

In all the above three WH cases, we found that both $ \Theta=0 $ and $ \sigma_{ab}=0 $, resulting in a gravitational temperature $T_{grav} = \dfrac{1}{2\pi}\vert\dot{u_{a}}z^{a}\vert $, i.e., the gravitational temperature is proportional to the acceleration. We recall the Tolman law to write $ \dfrac{dT}{T} \sim -a_{l} dx^{l} $, and consequently infer about the relationship between $ T_{0} $ and $ T_{grav} $. For the Ellis WH, the relationship becomes trivial, but for the other two WHs the relationship is non-trivial. The relation between the two temperatures for the Darmour-Solodukhin WH becomes: $ T_{0}=T_{grav}\left(\dfrac{2\pi \tilde{T}}{m}\right)\dfrac{r^2}{\sqrt{r-2m}}$, and for the exponential metric WH it is $T_{0}=T_{grav} \left(\dfrac{2\pi\tilde{T}}{m}\right) r^2 e^{2m/r}$. It is clear that the two temperatures are somehow related to each other as the gravitational temperature is proportional to the acceleration, whereas in the Tolman case, the temperature gradient is proportional to the acceleration. For a fixed $ r $, if one temperature is known, then the another one can be computed from these relations.

\section{Summary}
In this paper we have analyzed the gravitational entropy of traversable wormholes from the perspective of two different proposals, and have tried to make a comparison between the two proposals in terms of their applicability to the different wormhole geometries. We considered a variety of traversable WHs in order to analyze thoroughly the behaviour of GE in such cases, and determine whether the GE proposals considered by us can provide us with a viable measure of GE. Indeed, we found that the GE proposals do give us a consistent measure of GE in most of them.

First we adopted a phenomenological approach of determining the gravitational entropy of traversable wormholes \cite{entropy1,entropy2}. Secondly, we have also checked the result of applying the CET proposal \cite{CET} on traversable wormhole geometries of Petrov type D (excepting one above and one in the Appendices). It is clear that although the CET proposal is very successful in various astrophysical and cosmological cases, the definition of temperature given by CET  fails in the case of Ellis wormhole. In a more cosmologically significant case, i.e., traversable wormholes that mimic black holes, namely the Darmour-Solodukin wormhole, we find that it possesses a viable gravitational entropy for both the Weyl scalar proposal and the CET proposal, with some differences between them.

Although both the Rudjord and the CET proposals were successful in the case of the exponential metric wormhole, their behavior differed which was expected as the former was a purely geometric proposal whereas the latter one is based on relativistic thermodynamics. The Rudjord proposal is directly inspired from the fact that black hole entropy is related to its geometry, but the CET proposal considers a much more local view of the system. In the case of NUT wormholes, both the WHs were checked thoroughly using the Rudjord method and additional changes were made according to \cite{entropy2} both in the magnitude of $ P_{i} $ and in the additional vector component contributions. For the case of the first NUT wormhole the definition of $ P_{1} $ is not suitable as it gives multiple discontinuities at different $ r $ for different combination of parameters, and it did not improve even after introducing the additional angular contribution of $ \theta $ in the definition of the gravitational entropy density. In the appendix, the case for the traversable AdS wormhole, which is also a NUT wormhole, the $ P_{1} $ definition is not so bad as in the previous one. The overall distribution of the gravitational entropy density changed significantly by tilting in a direction if we introduce additional angular contribution, but for $ P_{2} $ the behaviour is regular and it only changes shape because the angular component brings in additional entropy into the system. Overall, for wormholes where $ g_{t\phi} $ is nonzero, the definition using $ P_{2} $ and both the vector directional components gives us a more robust result. Below we are presenting our results in a systematic manner.

\begin{enumerate}
\item For the Ellis wormhole we got a well-behaved gravitational entropy density using the Weyl scalar (Rudjord) proposal. However for the CET proposal, though we have a non zero gravitational energy density, but the gravitational temperature becomes zero thereby making it impossible to calculate the GE. As the definition of the gravitational temperature is \emph{ad hoc} in the CET proposal, we have introduced a new gravitational temperature to obtain the GE, which we found to be well-behaved. The GE density in the Weyl scalar proposal is always vanishing at the throat of the WH, and the radial variation rate for the CET GE is always finite at the throat.

\item For the Darmour-Solodukhin wormhole, both the Weyl scalar proposal and the CET proposal gives us a finite viable measure of gravitational entropy. The rate of radial variation of GE and the local piecewise CET GE blows up at the throat of the WH.
\item For the exponential metric wormhole too, we obtain viable measures of GE for both the Weyl scalar and the CET proposals. In the Weyl scalar case, the GE density peaks near the throat on the other side, and the CET rate of radial variation of GE decreases and goes to zero, crossing the throat on the other side of the universe. 
\item For the NUT wormhole we have computed the gravitational entropy density using both the functions $ P_{1} $ and $ P_{2} $, taking into account both the radial and angular components in the definition of $ s $. We have shown explicitly that for these wormholes, $ P_{2} $ gives us a viable measure of GE density. In this case, the CET proposal cannot provide us with an unique expression for gravitational entropy as the metric is not strictly of Petrov type D.
\end{enumerate}
Of the four WHs considered above, three of them are static spherically symmetric cases which are in equilibrium (i.e., the Ellis WH, the DS WH, and the exponential metric WH). Therefore, we applied the famous Tolman law on these spacetimes,  which says that thermal equilibrium can exist within a temperature gradient if a gravitational field is present there. We determined the Tolman temperature, and compared it to the temperature of the gravitational field.

In the Appendix section we have included the analysis for the NUT wormhole in the AdS spacetime as an additional case study. In this case, the Weyl proposal is applied for both the definitions of $ P_{1} $ and $ P_{2} $. In both these cases, the radial and the angular contributions together gives us a complete picture, but as these metrics have a nonzero $ g_{t\phi} $ component, we want to make a point that $ P_{2} $ (with both the radial and angular contributions) gives us a far more complete and viable measure of gravitational entropy density. Another traversable wormhole system proposed by Maldacena et al have been analyzed to examine the gravitational entropy behavior of a traversable wormhole connecting two black holes. In the Maldacena wormhole ansatz, both the Weyl proposal and the CET proposal gives us zero gravitational entropy density, making the system nonphysical from the thermodynamic perspective of gravitational entropy. For the sake of completeness we have analyzed the extremal magnetic BHs of this system and showed that the relevant functions, i.e., the gravitational energy density, the gravitational temperature, the ratio of curvature scalars and the gravitational energy density to gravitational temperature are continuous in the BH and WH junction. An important thing to note here is that the gravitational entropy of the extremal magnetically charged black holes on a horizon conforms with the Hawking-Bekenstein entropy of a black hole.

\section{Concluding remarks}

In this paper we have shown that the behavior of the gravitational entropy density function of a system depends on the definition employed to compute its value, and varies on a case by case basis. The two proposals which have been compared in this article are the Weyl scalar proposal and the CET proposal. In some cases the pure geometric method, i.e., the Weyl scalar proposal, provides us a good picture for the ratio of curvature scalars, $ P_{1} $. In the case of wormholes which have a nonzero $ g_{t\phi} $ term, the pure Weyl square $ P_{2} $ seems to work better. It is also important to consider both the radial and angular contributions in the definition of gravitational entropy. On the other hand, the CET proposal provides an unique gravitational entropy only for the Petrov type D and N spacetimes, but yields a far more nuanced result as it originates from the relativistic thermodynamic considerations. The definition of gravitational temperature proposed in the CET paper \cite{CET} does not give us an acceptable value in some of the cases. It has already been mentioned that the CET proposal is not dependent on the definition of temperature. Therefore a new definition of gravitational temperature can be used to suit the purpose. An important point which one must remember in this context is that, here we are considering only the wormhole geometry although the matter source may differ in each case. As we are interested in the gravitational entropy, the nature of matter source for the wormholes does not affect our analysis. Another important point to be noted is that the CET proposal have been applied to wormhole systems for the first time, and it is interesting to note its difference as compared to the Rudjord proposal.

We want to reiterate that by being frame-independent, the Weyl scalar proposals lost their connection
with the worldlines of physical fluids, and for this reason we cannot use these proposals to examine the GE for an important class of observers that traverse a WH, like the one we considered for the Ellis WH using the CET proposal. Since in a spherically symmetric static frame, the 4-velocity is aligned along the time axis, the only straightforward interpretation of the expressions in our calculations (obtained for the Weyl scalar proposals) is the variation of GE density along the radial direction, defined in this frame. These calculations demonstrate a GE that yields a different constant for each 2-sphere labelled by specific values of the coordinate $ r $. Other criticisms on the Weyl scalar proposals are available in \cite{CET}.

We argue that for any traversable wormhole to exist, it must have a viable gravitational entropy. So it is important to compare and study the proposals of gravitational entropy in the context of traversable wormholes. Obviously, the CET proposal of gravitational entropy is much more physically important, as it takes into account the restrictions imposed by relativistic thermodynamics. Not only is the interpretation of the gravitational energy density established on a firm footing, but it also provides the freedom to choose the gravitational temperature.
For spacetimes not belonging to Petrov types D and N,  several algebraic decompositions of the Bel-Robinson tensor can be computed, and the resulting expressions for CET gravitational entropy can also be studied. The CET proposal depends on frames, and this frame-dependence is a very important property that allows a link with thermodynamics. Though the Weyl scalar proposals does not contain the thermodynamic aspects in it, they can provide a good viable measure of GE for cases where the CET and other proposals may not. So we want to emphasize that a given proposal might work for some spacetimes and not for others. However, all spacetimes are not equally interesting or valid from a physical point of view. Therefore the GE proposals which work for the most physically meaningful spacetimes are the most interesting and relevant ones. From this view point, the connection to thermodynamics and dependence on frames makes the CET proposal more physically viable. Hence it appears that it will indeed be challenging to develop an universal proposal of gravitational entropy.

Another important point to note here is that we are not checking whether the entropy is increasing with time but whether these proposals can provide us with a viable expression of GE, as for the equilibrium states of the static cases the condition $ u^{a}\partial_{a}S_{grav}=0 $ holds. However, the thermodynamic study of equilibrium states for such self-gravitating systems are very interesting, like the study of Antonov instability in relativistic systems \cite{paddy,roupas} can important in these cases. Such studies of equilibrium states of these self-gravitating systems may provide us with new information.

In conclusion, from our analysis and the corresponding plots, it appears that for the traversable wormholes the GE density function will be well-defined if the definition of the vector field $ \mathbf{\Psi} $ is modified, either in the magnitude ($ P $), or in its direction (having additional angular components). A similar feature was observed in our earlier work \cite{GC} in the case of accelerating black holes. For the CET proposal, a new definition of gravitational temperature can be used in order to avoid the appearance of null gravitational temperature. All the static cases produced well defined unique GE in both the Weyl scalar and the CET GE proposals, with minor differences, indicating that the concept of gravitational entropy although new and contentious, does have some sense of theoretical robustness.

\section*{Data Availability}
Data sharing is not applicable to this article as no datasets were generated or analysed during the current study. All figures were plotted with Maple software using the theoretical equations.

\section*{Acknowledgments}
We are thankful to the anonymous reviewers for their comments and constructive suggestions. SC is grateful to CSIR, Government of India for providing junior research fellowship. SG gratefully acknowledges IUCAA, India for an associateship and CSIR, Government of India for approving the major research project No. 03(1446)/18/EMR-II. RG thanks National Research Foundation, South Africa, for research support.

\section*{Appendix: SOME ADDITIONAL NOTES}
In this section we will briefly analyse the case of two more traversable wormholes. These are additional case studies to support our conclusions, but less significant from the point of view of the observed features.

\subsection*{Appendix-I: Traversable AdS wormhole}

The following metric represents a stationary NUT wormhole with a negative cosmological constant with a nonlinear sigma model as source, which was  illustrated in \cite{WH2}:
\begin{equation}
ds^2=dz^2+\dfrac{\rho^2(z)}{4}\left[ -Q^2(d\tau+cos\theta d\phi)^2 +(d\theta^2+sin^2\theta d\phi^2) \right],
\end{equation}
where $ \rho(z)=\sqrt{\dfrac{3(K-8)}{4\vert\Lambda\vert}} \cosh\left(\dfrac{\vert\Lambda\vert^{1/2}}{\sqrt{3}}z\right), $ and $ Q^2= \dfrac{K}{4}.$
In addition to this, $ \Lambda<0 $ if the system has to satisfy the Einstein equations. Here $ Q $ is the NUT parameter and must be an even integer for the solution to be single-valued \cite{GP}. The asymptotic NUT-AdS regions with $ z\rightarrow \pm \infty $ are connected by this wormhole at the throat $ z=0 $. This spacetime has no curvature singularities and is locally regular, and is therefore an object of our interest in this paper.

The determinant of the metric for the traversable AdS wormhole is given by the following:
\\
\begin{equation}
g=-{\frac {27\, \left( K-8 \right) ^{3} \left( \cosh \left( 1/3\,\sqrt {
 \left| \Lambda \right| }\sqrt {3}z \right)  \right) ^{6}{K}^{2}
 \left( \sin \left( \theta \right)  \right) ^{2}}{65536\, \left(
 \left| \Lambda \right|  \right) ^{3}}}.
\end{equation}
Consequently the Weyl curvature scalar square is given by the following:
\begin{equation}
 W={\frac {4\, \left( {K}^{2}+16 \right) ^{2}{\Lambda}^{2}}{27\, \left(
\cosh \left( 1/3\,\sqrt { \left| \Lambda \right| }\sqrt {3}z \right)
 \right) ^{4} \left( K-8 \right) ^{2}}}.
\end{equation}
The Kretschmann curvature scalar ($ \tilde{K} $) for the AdS wormhole is given by the equation:
\begin{align}
\left. \tilde{K}\right.&= \frac {11\,{\Lambda}^{2}}{36\, \left( \cosh \left( 1/3\,\sqrt {
 \left| \Lambda \right| }\sqrt {3}z \right)  \right) ^{4} \left( K-8
 \right) ^{2}} \Bigg[ {\frac {96\, \left( \cosh \left( 1/3\,\sqrt {
 \left| \Lambda \right| }\sqrt {3}z \right)  \right) ^{4} \left( K-8
 \right) ^{2}}{11}}-\nonumber\\
 &{\frac { \left( 8\,K-64 \right)  \left( {K}^{2}+12
\,K-32 \right)  \left( \cosh \left( 1/3\,\sqrt { \left| \Lambda
 \right| }\sqrt {3}z \right)  \right) ^{2}}{11}}+{K}^{4}+ {\frac {8\,{K
}^{3}}{11}}+{\frac {368\,{K}^{2}}{11}}-{\frac {256\,K}{11}}+{\frac {
3072}{11}} \Bigg].
\end{align}
\begin{figure}[ht]
    \centering
    \subfloat[Subfigure 1 list of figures text][]
        {
        \includegraphics[width=0.48\textwidth, height=0.155\textheight]{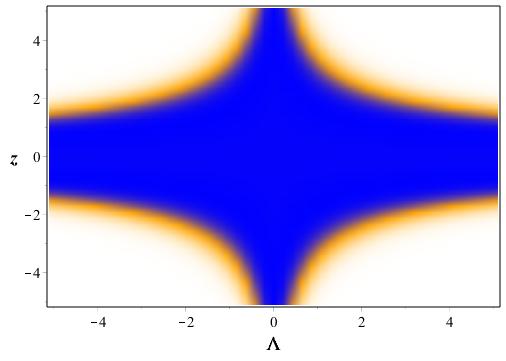}
        \label{fig:subfig3d}
        }
    \subfloat[Subfigure 2 list of figures text][]
        {
        \includegraphics[width=0.48\textwidth, height=0.157\textheight]{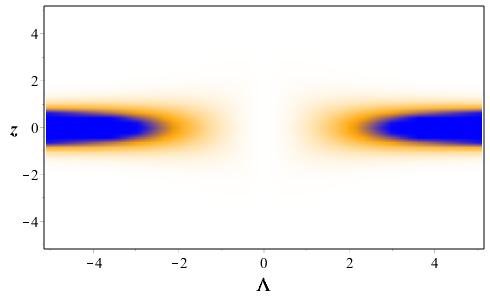}
        \label{fig:subfig4d}
        }
    \caption{(a)Variation of $ P_{1}^2 $ for traversable AdS wormhole with $ z $ and the parameter $\Lambda$. (b)Variation of $ P_{2} $ for traversable AdS wormhole with $ z $ and the parameter $\Lambda$. In both the cases the NUT parameter is $K = 16$. Here the blue region represents high positive values and it gradually decreases through the yellow region to the white colored region.}
   \label{AdSpp}
\end{figure}
\begin{enumerate}
\item \textbf{Weyl scalar proposal:} From the above expressions, the ratio of the two curvature scalars is obtained along straightforward calculations, and is given below:
\begin{align}\label{AdSp1}
\left. P_{1}\right.&= \frac {4\,\sqrt {33}}{33}\Bigg[ \left( {K}^{2}+16 \right) ^{2}
 \Bigg( {\frac {96\, \left( \cosh \left( 1/3\,\sqrt { \left| \Lambda
 \right| }\sqrt {3}z \right)  \right) ^{4} \left( K-8 \right) ^{2}}{11}}  \nonumber \\
  & - {\frac { \left( 8\,K-64 \right)  \left( {K}^{2}+12\,K-32 \right)
 \left( \cosh \left( 1/3\,\sqrt { \left| \Lambda \right| }\sqrt {3}z
 \right)  \right) ^{2}}{11}} + {K}^{4}+{\frac {8\,{K}^{3}}{11}}+{\frac {368\,{K}^{2}}{11}}-{\frac {256\,K}{11}}+{\frac {3072}{11}} \Bigg) ^{-
1}\Bigg]^{\frac{1}{2}}.
\end{align}
The ratio of curvature scalars, i.e. $P_{1}^2$, is given in the expression \eqref{AdSp1}. For the sake of clarity, it is also illustrated graphically in FIG.\ref{AdSpp}(a).

\begin{figure}
\centering
\subfloat[]{\label{main:a1}\includegraphics[width=0.6\textwidth]{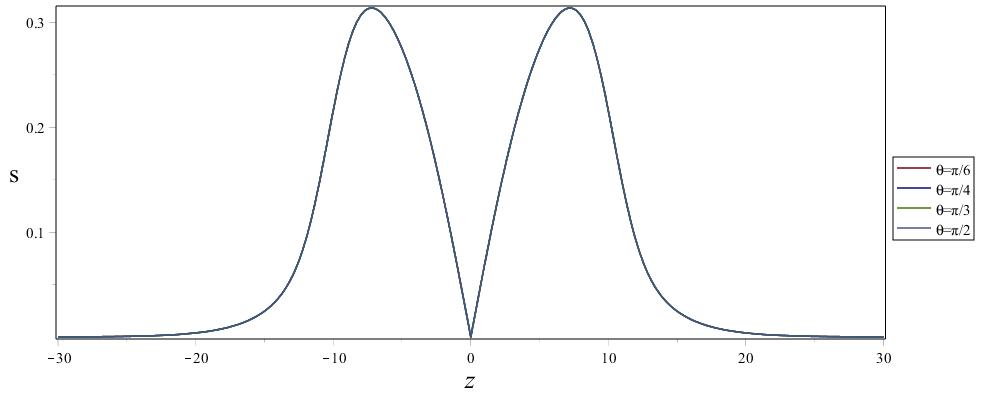}}
\par\medskip
\centering
\subfloat[]{\label{main:b1}\includegraphics[width=0.6\textwidth]{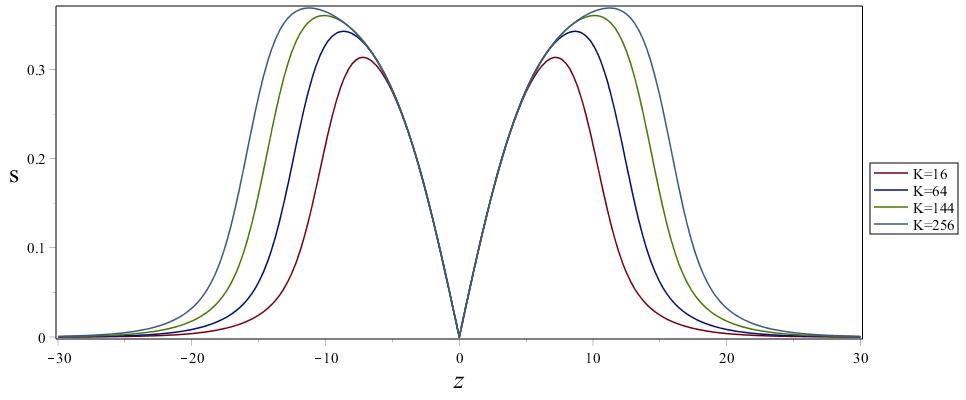}}
\par\medskip
\centering
\subfloat[]{\label{main:c1}\includegraphics[width=0.7\textwidth]{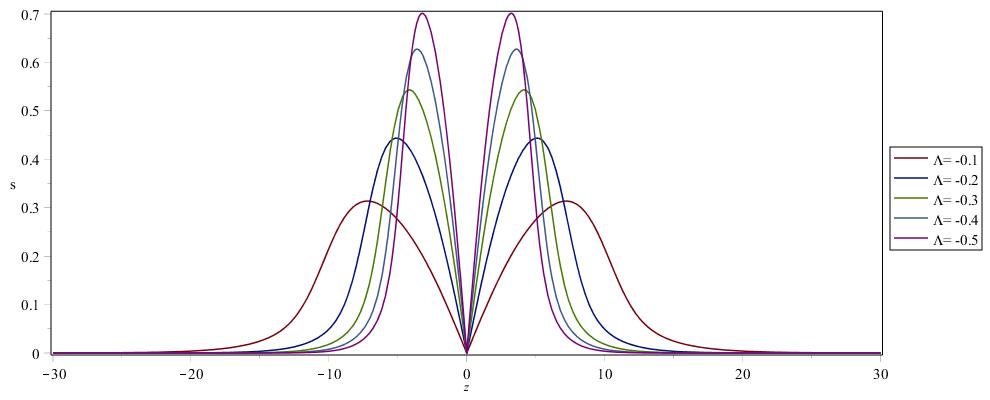}}

\caption{Variation of entropy density $ s $ of AdS traversable wormhole with $ z $ for different parameters. Here we have considered only the radial contribution as defined in \eqref{sd1}. Also the definition of gravitational entropy density is calculated using $ P_{1} $.}
\label{ads1z}
\end{figure}

\begin{figure}

\centering
\subfloat[]{\label{main:a2}\includegraphics[width=0.6\textwidth]{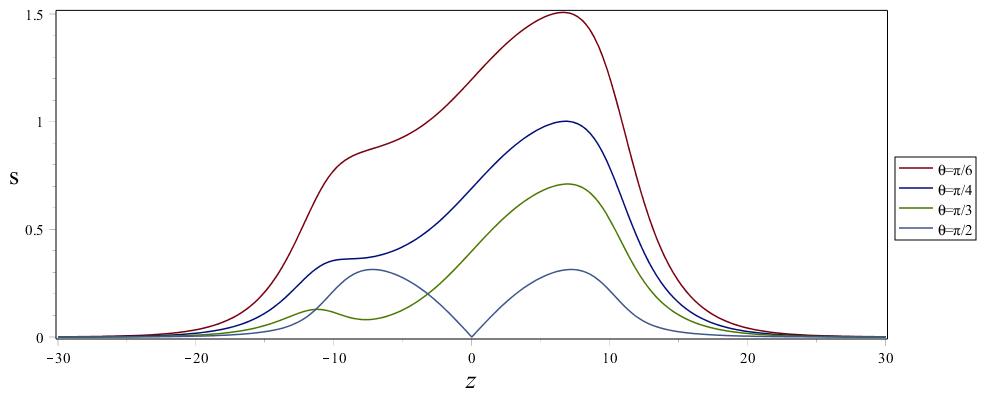}}
\par\medskip
\centering
\subfloat[]{\label{main:b2}\includegraphics[width=0.6\textwidth]{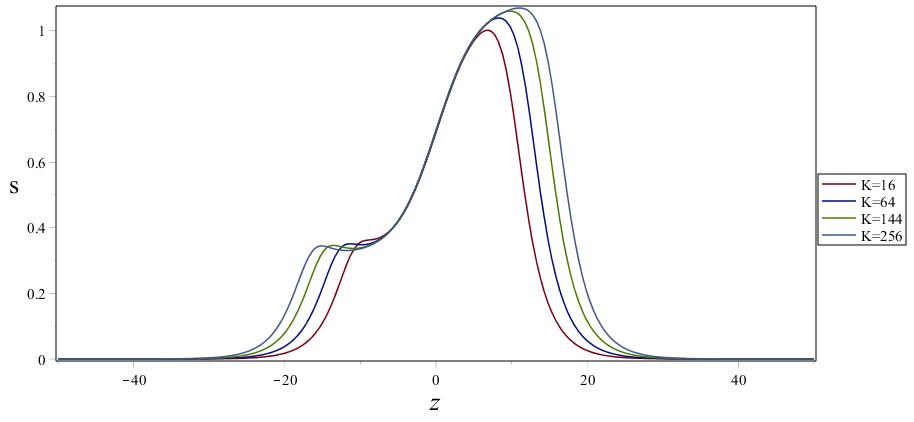}}
\par\medskip
\centering
\subfloat[]{\label{main:c2}\includegraphics[width=0.7\textwidth]{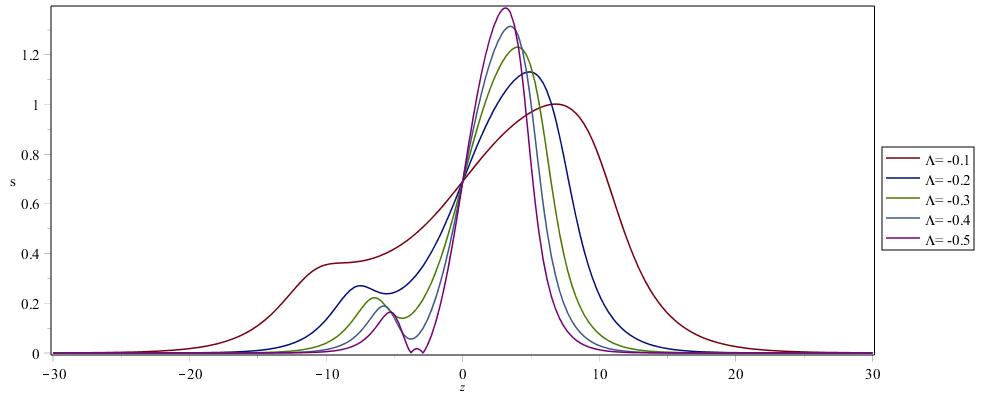}}

\caption{Variation of entropy density $ s $ of AdS traversable wormhole with $ z $ for  different parameters. We have considered both the radial and angular contribution \eqref{sd2} and the gravitational entropy density is calculated using $ P_{1} $.}
\label{ads1zt}
\end{figure}

In FIG.\ref{ads1z} and FIG. \ref{ads1zt} we have shown the variation of gravitational entropy density with different parameters. In both these cases, $ P_{1} $ is being used for the calculations. In FIG.\ref{ads1z} we have only taken the contribution of the radial component while determining the gravitational entropy density and in FIG. \ref{ads1zt} both the radial and angular contributions are taken into account.

In FIG.\ref{ads1z}(a) the variation of the gravitational entropy density of AdS wormhole is shown with angular orientation $\theta  $, where we have fixed the other parameters as $ \Lambda=-0.1, K=16 $. In this case we see no change as we did not include the angular contribution into our analysis. Next in FIG.\ref{ads1z}(b), the variation with the parameter $ K $ is being studied for $ \Lambda=-0.1, \theta=\dfrac{\pi}{4} $. As the value of $ K $ increases, so does the value of gravitational entropy density. In the last figure FIG.\ref{ads1z}(c), the variation of the gravitational entropy density of traversable AdS wormhole is shown with various negative values of cosmological constant $ \Lambda $ for the fixed parameter values as $ K=16, \theta=\frac{\pi}{4} $. Here we can clearly observe that with increasing negative value of the cosmological constant the peak value of the gravitational entropy density increases.
FIG. \ref{ads1zt} shows these variations with a higher sensitivity including both the radial and angular contributions in the entropy density. FIG. \ref{ads1zt}(a) shows the variation with the angular orientation while fixing the other parameters at $ \Lambda=-0.1, K=16 $. Here each angular orientation gives different gravitational entropy unlike the previous case. FIG. \ref{ads1zt}(b) gives us the nature of variation of entropy density with $ K $ when we fix $ \Lambda=-0.1, \theta=\dfrac{\pi}{4} $. In FIG. \ref{ads1zt}(c) the dependence of gravitational entropy density on the cosmological constant is being depicted, where we have chosen our free parameters as $ K=16, \theta=\frac{\pi}{4} $. Here the overall dependence remains the same but the extra contribution from the angular components in the gravitational entropy density makes it non zero at the throat region unlike in the previous case, which means that the gravitational entropy density is continuous through the wormhole throat.
 \begin{figure}
\centering
\subfloat[]{\label{main:a3}\includegraphics[width=0.6\textwidth]{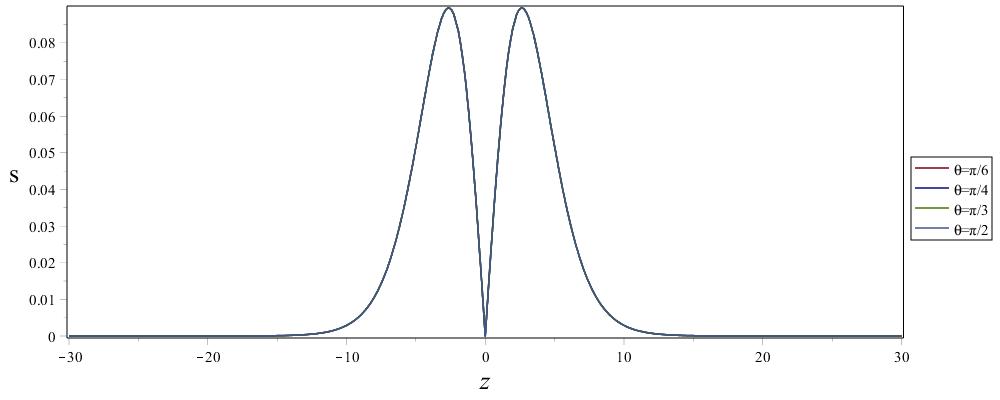}}
\par\medskip
\centering
\subfloat[]{\label{main:b3}\includegraphics[width=0.6\textwidth]{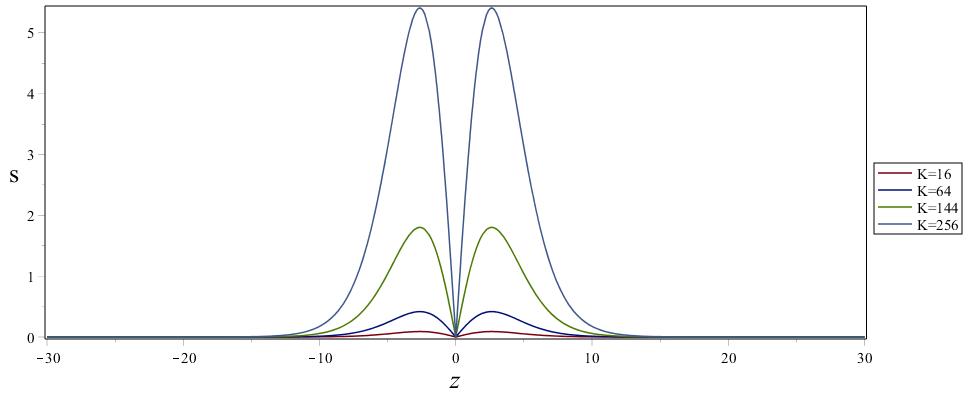}}
\par\medskip
\centering
\subfloat[]{\label{main:c3}\includegraphics[width=0.7\textwidth]{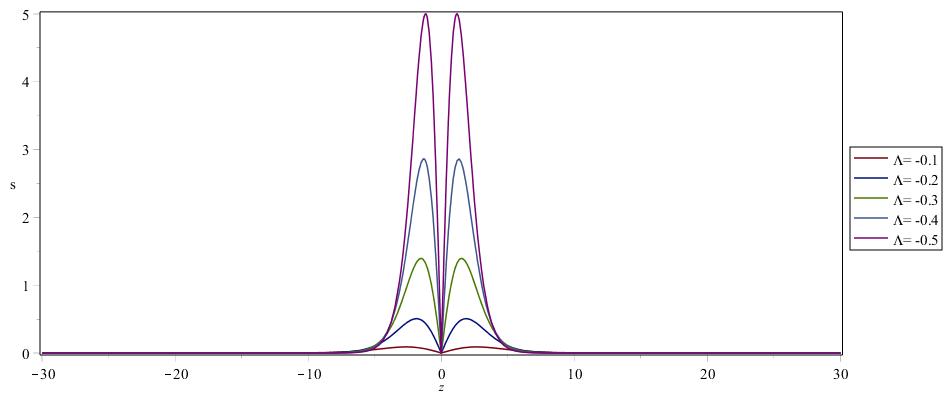}}

\caption{Variation of entropy density $ s $ of AdS traversable wormhole with $ z $ for  different parameters. Here we have considered only the radial contribution as in \eqref{sd1}. The gravitational entropy density is calculated using $ P_{2} $.}
\label{ads2z}
\end{figure}

\begin{figure}

\centering
\subfloat[]{\label{main:a4}\includegraphics[width=0.6\textwidth]{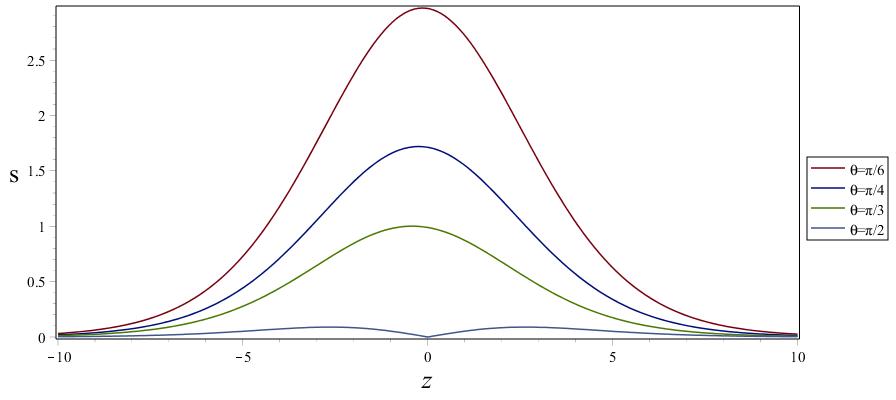}}
\par\medskip
\centering
\subfloat[]{\label{main:b4}\includegraphics[width=0.6\textwidth]{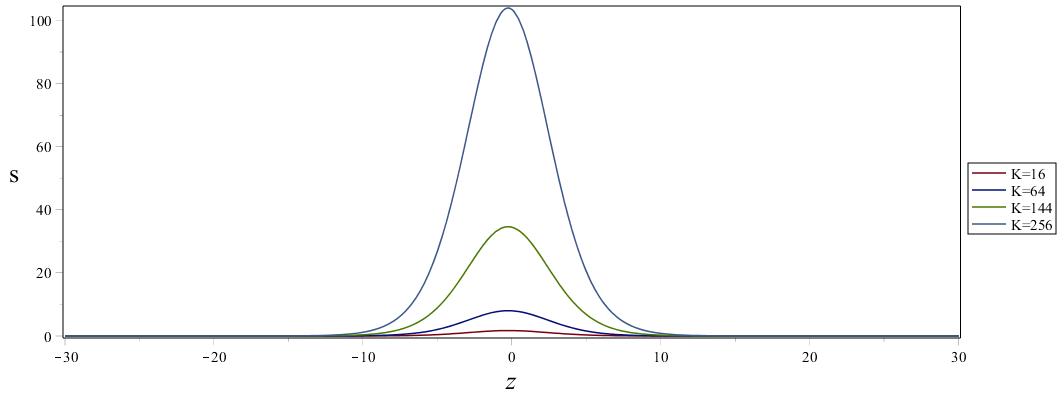}}
\par\medskip
\centering
\subfloat[]{\label{main:c4}\includegraphics[width=0.7\textwidth]{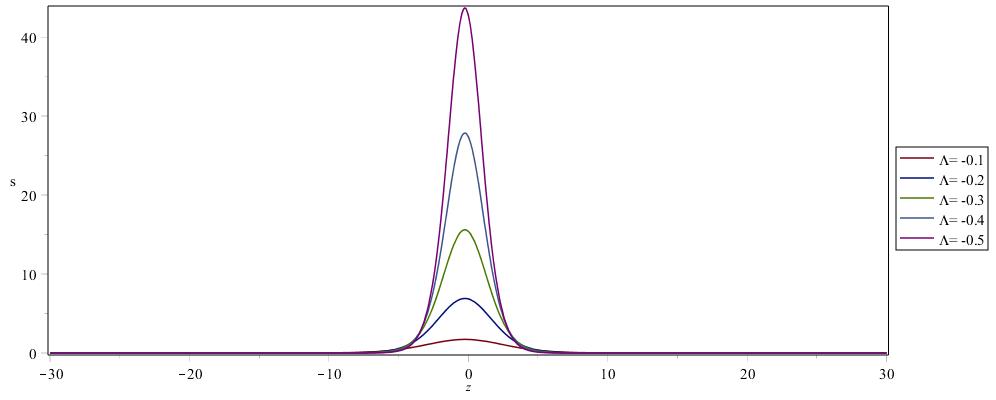}}

\caption{Variation of entropy density $ s $ of AdS traversable wormhole with $ z $ for  different parameters. Here we have considered both the radial and angular contribution \eqref{sd2}, and the gravitational entropy density is calculated using $ P_{2} $.}
\label{ads2zt}
\end{figure}

As we know that the $ g_{t\phi} $ component is also nonzero in the metric of AdS traversable wormhole, therefore the alternative definition of the gravitational entropy using $ P_{2} $ must be applied to see how the result differs from the former case. The graphical representation of $ P_{2} $ is shown in  FIG.\ref{AdSpp}(b).

Consequently in FIG. \ref{ads2z} and FIG. \ref{ads2zt} we have used $ P_{2} $ as the definition and used the radial contribution only to draw the graphs (in FIG. \ref{ads2z}), and in the later figures both the contributions of radial and angular components have been considered.

In FIG. \ref{ads2z}(a) we have chosen $ \Lambda=-0.1, K=16 $ to show the variation with the angular orientation while in FIG. \ref{ads2z}(b), the values $ \Lambda=-0.1, \theta=\dfrac{\pi}{4} $ are fixed to show the variation with the parameter $ K $. Similar to the case for $ P_{1} $, here too there are no changes in FIG. \ref{ads2z}(a) while the variations in FIG. \ref{ads2z}(b) are also similar to that of $ P_{1} $ except that the graphs are way more compact, i.e. the gravitational entropy density is localized in a much more smaller region when we consider $ P_{2} $. FIG. \ref{ads2z}(c) shows the variation of gravitational entropy density with the negative values of cosmological constant with the fixed parameters as $ K=16, \theta=\frac{\pi}{4} $ where we find that the magnitude of $ s $ increases with increasing negative values of the cosmological constant $ \Lambda $.

FIG. \ref{ads2zt}(a) shows the variation with $ \theta $ for $ \Lambda=-0.1, K=16 $. The introduction of the angular component changes the entropy density drastically for higher values of the angular orientation $ \theta $. If we fix $ \Lambda=-0.1, \theta=\dfrac{\pi}{4} $ as constants, then we obtain FIG. \ref{ads2zt}(b), which shows the variation with $ K $. Finally FIG. \ref{ads2zt}(c) shows the variation with $ \Lambda $ for the fixed parameter $ K=16, \theta=\frac{\pi}{4} $. The introduction of the angular component reduces the gravitational entropy density from a double peaked one to a single peaked function but the overall evenness with respect to the radius is not lost.

In general, the gravitational entropy density increases near the throat region when we consider both the radial and angular contribution for our analysis. Therefore we can say that the procedure involving the $ P_{2} $, which includes both the radial and angular contribution, is more suitable for the analysis of gravitational entropy in this case.
\end{enumerate}

\subsection*{Appendix-II: Maldacena wormhole ansatz}

Very recently Maldacena and Milekhin have discussed humanly traversable wormholes \cite{mal1}, where they have proposed a hypothetical connecting wormhole  between two oppositely charged magnetic blackholes. This is an interesting situation, worth analysing. The metric is given by
\begin{equation}\label{mwh}
ds^2=r_{e}^2[-(\rho^2+1)d\tau^2+\dfrac{d\rho^2}{(\rho^2+1)}+(d\theta^2+\sin^2\theta d\phi^2)],  \qquad \,\,\,\,\, -\rho_{c}\leq\rho\leq\rho_{c}
\end{equation}
\begin{figure}[ht]
    \centering
\includegraphics[width=0.5\textwidth]{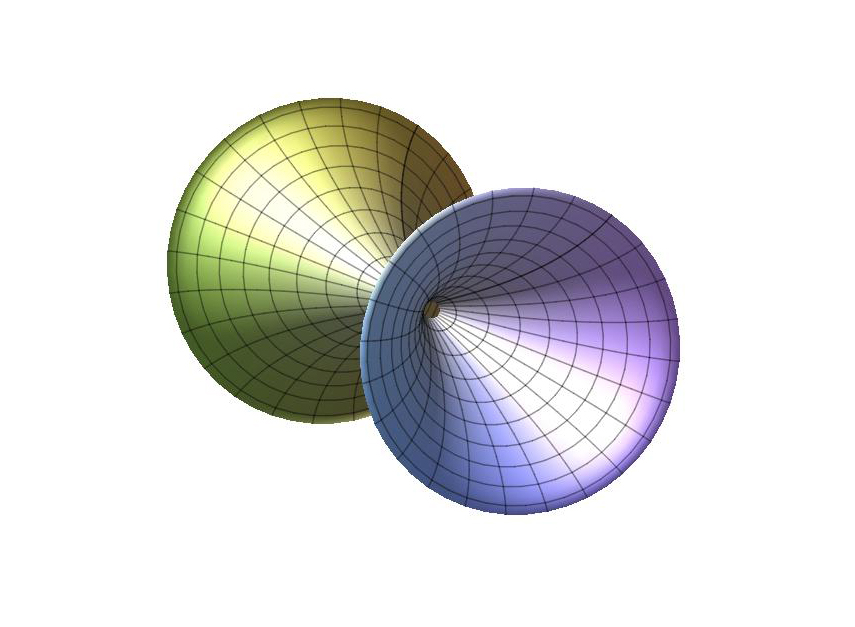}
\caption{Embedding diagram of Maldacena wormhole ansatz}
\label{Mal1}
\end{figure}
where $ \rho=\dfrac{l(r-r_{e})}{r_{e}^2} $ for $ \rho>>1 $, $ l=t/\tau $ and $ r-r_{e}<<r_{e} $ are the interrelations connecting the coordinates of \eqref{mwh} and \eqref{mbh}. Beyond the limit of $ \rho=\pm \rho_{c} $ of the wormhole region, the geometry is of two extremal magnetic blackholes. In Fig.\ref{Mal1} the wormhole is connecting two extremal magnetically charged black holes given by the following black hole metric \eqref{mbh}:

\begin{equation}\label{mbh}
ds^2=-fdt^2+\dfrac{dr^2}{f}+r^2(d\theta^2+sin^2\theta d\phi^2),
\end{equation}
where $ A=\dfrac{q}{2}cos\theta d\phi $; \,\,\,$\ell_{p}\equiv \sqrt{G_{4}}  $;\,\,\,$ r_{e}\equiv\dfrac{\sqrt{\pi}q\ell_{p}}{g_{4}} $; \,\,\,$ M_{e}=\dfrac{r_{e}}{G_{4}} $;\,\,\,$ f=\left(1-\frac{r_{e}}{r}\right)^2 $. Here $ q $ is the magnetic charge which is an integer and $M_{e}$ is the mass of the BH at extremality. In the near horizon region, $ r_{e} $ sets the radius of curvature and also the size of the 2-sphere. At the extremal limit as $ r\rightarrow r_{e} $, an infinite throat develops. Using the principal null tetrads we can easily get the following value of the Weyl scalar:

$$ \Psi_{2}=\dfrac{r_{e}(r_{e}-r)}{r^4} .$$
Using the CET proposal we obtain the gravitational energy density of the BH as:
\begin{equation}\label{mr}
\rho_{grav}=\dfrac{\alpha}{4\pi}\left\vert\dfrac{r_{e}(r_{e}-r)}{r^4}\right\vert.
\end{equation}
Similarly the gravitational temperature can be computed as
\begin{equation}\label{mt}
T_{grav}=\dfrac{1}{2\pi}\left\vert\dfrac{r_{e}}{r^2}\right\vert .
\end{equation}
Finally the ratio of gravitational energy density to gravitational temperature for this BH is given by
\begin{equation}\label{ms}
\dfrac{\rho_{grav}}{T_{grav}}=\frac{\alpha}{2}\left\vert\dfrac{(-r+r_{e})}{r^2}\right\vert .
\end{equation}
For the sake of completeness, and to check the validity of CET proposal, we can compute the gravitational entropy of this BH on a surface with radius $ R $ as:
\begin{equation}
S_{grav}=\int_{0}^{R}\int_{0}^{\pi}\int_{0}^{2\pi}\frac{\alpha}{2}\left\vert\dfrac{(-r+r_{e})}{r^2}\right\vert \dfrac{r^2 \sin\theta}{(1-\frac{r_{e}}{r})}dr d\theta d\phi=\alpha \pi R^2 .
\end{equation}
Thus the CET gravitational entropy of the magnetized extremal BH is directly proportional to the horizon area, conforming with the definition of Hawking-Bekenstein entropy.
Moreover, the ratio for the curvature scalars for the BH is given by the following expression:
\begin{equation}\label{mp}
P_{1}^2=\dfrac{6(-r+r_{e})^2}{7r_{e}^2-12r_{e}r+6r^2} .
\end{equation}

Next we calculate the Weyl scalar proposal gravitational entropy density for the connecting wormhole, which turns out to be:
$s=k_{s}\vert\nabla.\Psi\vert=0$.
Surprisingly, we find that the gravitational entropy of such wormholes vanishes in this proposal. This shows that either the proposal itself is not valid in this case, or the wormhole itself is nonphysical in nature.

Similarly for the CET proposal of gravitational entropy the gravitational energy density of the connecting wormhole is found to vanish, raising questions on the physical viability of such WHs. The relevant expressions are given below:
\begin{equation}
\rho_{grav}=0; \,\,\,\, T_{grav}=\dfrac{1}{2\pi}\left\vert  \dfrac{\rho}{r_{e}\sqrt{1+\rho^2}}   \right\vert;\,\,\,\,\, s_{grav}=0 \,\,(\rho\neq 0).
\end{equation}
As the gravitational entropy is zero even in the CET proposal, we can say that it conforms to our analysis of the extremal magnetic BHs. As $ r\rightarrow r_{e} $, the ratio of curvature scalars in \eqref{mp} reduces to zero, indicating a zero gravitational entropy density for the Weyl proposal. In this limit, for the CET proposal the gravitational energy density in \eqref{mr} also reduces to zero and the temperature in \eqref{mt} becomes $ \sim \left\vert\dfrac{1}{2\pi r_{e}} \right\vert$, which matches with the wormhole gravitational temperature in the limit $ \rho>>1 $. Consequently the ratio in \eqref{ms} becomes zero in this limit, thereby matching with the wormhole counterpart.
The interesting point to note here is that the gravitational temperature is nonzero, but at $ \rho=0 $. Thus at $ \rho=0 $ we may still have a finite entropy, as in that case the ratio of gravitational energy density to gravitational temperature assumes the form $ \frac{0}{0} $, but these proposals are not yet equipped to address such cases.

\end{document}